\documentclass[11pt,letterpaper]{article}

\usepackage{times}  
\usepackage{booktabs}
\usepackage{graphicx}
\usepackage{url}
\usepackage{color}
\usepackage{multirow}
\usepackage{amsmath}
\usepackage{amssymb}
\usepackage{amsthm}
\usepackage{cite}  
\usepackage{xspace}
\usepackage{array}
\usepackage{enumitem}
\usepackage{subfigure}
\usepackage{epstopdf}
\usepackage[normalem]{ulem}
\usepackage{microtype}
\DisableLigatures{encoding=*, family=*}

\renewcommand{\paragraph}[1]{\smallskip\noindent {\bf #1}}

\newtheorem{theorem}{ {\noindent{\bf Theorem}}}

\begin{document}

\title{Optimal Repair Layering for Erasure-Coded Data Centers: From Theory to
Practice}

\author{Yuchong Hu$^1$, Xiaolu Li$^2$, Mi Zhang$^2$, Patrick P. C. Lee$^2$,
Xiaoyang Zhang$^1$, Pan Zhou$^1$, Dan Feng$^1$\\
$^1$Huazhong University of Science and Technology\\
$^2$The Chinese University of Hong Kong}

%
%
%
%
%

\date{}

\maketitle

\begin{abstract}
Repair performance in hierarchical data centers is often bottlenecked by
cross-rack network transfer.  Recent theoretical results show that the
cross-rack repair traffic can be minimized through repair layering, whose
idea is to partition a repair operation into inner-rack and cross-rack layers.
However, how repair layering should be implemented and deployed in practice
remains an open issue.  In this paper, we address this issue by proposing a
practical repair layering framework called DoubleR.  We design two families of
practical double regenerating codes (DRC), which not only minimize the
cross-rack repair traffic, but also have several practical properties that
improve state-of-the-art regenerating codes.  We implement and deploy DoubleR
atop Hadoop Distributed File System (HDFS), and show that DoubleR maintains
the theoretical guarantees of DRC and improves the repair performance of
regenerating codes in both node recovery and degraded read operations. 
\end{abstract}

\section{Introduction}
\label{sec:introduction}

As data center storage expands at scale, failures are more prevalent in
storage subsystems \cite{ford10,schroeder07,jiang08}.  To maintain data
availability and durability at low cost, modern data centers increasingly
adopt {\em erasure coding} to protect data storage with a significantly low
degree of redundancy, while still preserving the same fault tolerance as
traditional replication.  
At a high level, an erasure code works by transforming a set of uncoded
fixed-size units, called {\em blocks}, into a larger set of coded blocks, such
that the set of uncoded blocks can be reconstructed from any subset of the
same number of coded blocks.  Each set of coded blocks is called a 
{\em stripe}, and a data center stores multiple stripes that are independently
erasure-coded.  By distributing the coded blocks (or blocks in short) of each
stripe across distinct storage nodes, a data center can provide fault
tolerance against node failures. 
Field studies have shown the benefits of erasure coding in saving storage
overhead in production.  For example, Azure \cite{huang12} and Facebook
\cite{muralidhar14} reportedly reduce storage redundancy to 1.33$\times$ and
1.4$\times$, respectively, as opposed to 3$\times$ in traditional triple
replication \cite{ghemawat03,calder11}.

A drawback of erasure coding is the high repair cost.  Repair operations are
triggered when issuing degraded reads to unavailable blocks, or recovering
lost blocks from node crashes.  In both cases, repairing each failed block in
erasure-coded storage must retrieve multiple available blocks from other nodes
for reconstruction.  This leads to substantial {\em repair traffic}, defined
as the amount of data transferred for repair.  Facebook \cite{rashmi13}
reports that its erasure-coded data center generates a median size of 180
terabytes of repair traffic per day, which in turn limits the bandwidth
resources available for foreground jobs.  In practice, bandwidth resources
available for repair tasks are often throttled \cite{huang12,silberstein14} to
limit their adverse impact on other application traffic. 
Thus, there has been an extensive literature on mitigating the repair cost
(see \S\ref{sec:related}).  In particular, {\em regenerating codes}
\cite{dimakis10} are a special class of erasure codes that provably minimize
the repair traffic, and there are many follow-up theoretical studies on
regenerating codes.  In addition, recent studies (e.g.,
\cite{chen14,li15,rashmi15,pamies16}) have prototyped regenerating codes and
evaluated their practical performance in networked environments.

However, regenerating codes are still limited in addressing the
{\em hierarchical} nature of data centers.  Modern data centers organize nodes
in racks, and are {\em oversubscribed} to control operational costs
\cite{cisco04}.  Although full-bisection bandwidth is available within a rack,
cross-rack bandwidth is constrained.  Typical oversubscription ratios range
from 5:1 to 20:1 \cite{ahmad14,benson10,vahdat10} (i.e., the available
cross-rack bandwidth per node is only 1/5 to 1/20 of the inner-rack bandwidth
in the worst case); in some extremes, the ratio could reach 240:1
\cite{greenberg09}.   Cross-rack links are also shared by replica writes
\cite{chowdhury13} or shuffle/join traffic of computing jobs
\cite{ahmad14,jalaparti15}.  Note that geo-distributed data centers
\cite{ford10,aguilera13} also exhibit the similar hierarchical nature, as the
bandwidth resources across geographical regions are limited \cite{aguilera13}
and intermittently congested \cite{cho12} as opposed to within the same region.

To maximize fault tolerance, existing erasure-coded data centers often place
each block of a stripe in a distinct node that resides in a distinct rack
(i.e., one block per rack)
\cite{ford10,huang12,sathiamoorthy13,muralidhar14,rashmi14,rashmi15}.  We call
this {\em flat block placement}, which allows a data center to tolerate the
same numbers of node failures and rack failures.  However, this inevitably
makes the repair of any failed block retrieve available blocks from other
racks, and hence incurs substantial {\em cross-rack} repair traffic, even
though the repair traffic can be minimized by regenerating codes.

In this paper, we propose {\em DoubleR}, a repair framework that is designed
to minimize the cross-rack repair traffic for hierarchical data centers.
DoubleR advocates a concept called {\em repair layering}, which splits a
repair operation into inner-rack and cross-rack layers and trades (abundant)
inner-rack bandwidth for (constrained) cross-rack bandwidth.  Specifically,
DoubleR opts for {\em hierarchical block placement}, which places multiple
blocks of a stripe per rack, so as to minimize the cross-rack repair traffic
at the expense of reducing rack-level fault tolerance.
To repair a failed block, one selected node in each rack can perform partial
repair operations internally using the available blocks from the same rack. It
then sends partially repaired results across racks to a destination node,
which combines the partially repaired results from multiple racks to
reconstruct the failed block.  Through repair layering, it is theoretically
proven that the cross-rack repair traffic can be minimized through a new class
of regenerating codes called {\em double regenerating codes (DRC)}
\cite{hu16}.  We augment the theoretical results in \cite{hu16} into the
repair framework DoubleR, and make the following contributions from an applied
perspective:
\begin{itemize}[leftmargin=*] \itemsep=0pt \parskip=0pt
\item
We provide numerical analysis to show that repair layering significantly
reduces the cross-rack repair traffic compared to state-of-the-art
regenerating codes.  We also provide reliability analysis to study the
trade-off between the minimized cross-rack repair traffic and the reduced
rack-level fault tolerance.  While similar numerical and reliability analysis
on erasure-coded storage has been found in the literature (e.g.,
\cite{huang12,sathiamoorthy13,li15}), our analysis is new by specifically
addressing the hierarchical nature of data centers. 
\item
We propose two families of constructions of double regenerating codes (DRC).
Both constructions preserve the theoretical guarantees of minimizing the
cross-rack repair traffic, and are specifically designed for the practical
deployment in real-world data centers.
\item
We implement a DoubleR prototype atop Facebook's HDFS \cite{hadoop20}.  We
extensively parallelize the operations to mitigate repair overhead.  We also
export APIs that can incorporate not only DRC but also existing regenerating
codes.
\item
We conduct testbed experiments on evaluating different erasure codes using our
DoubleR prototype.  We show that DRC increases the single failed node recovery
throughput and reduces the degraded read time to an unavailable block.  Our
results also conform to the numerical results of bandwidth savings of DRC. 
\end{itemize}

The remainder of the paper proceeds as follows. 
In \S\ref{sec:overview}, we introduce and motivate the design of DoubleR. 
In \S\ref{sec:drc}, we present analytical results on DRC.
In \S\ref{sec:practical}, we propose two practical constructions of DRC.  
In \S\ref{sec:hadoop}, we describe the implementation details of DoubleR. 
In \S\ref{sec:experiments}, we present experimental results.
In \S\ref{sec:related}, we review related work.
In \S\ref{sec:discuss}, we provide a detailed discussion on the design
trade-offs of DRC, and finally
in \S\ref{sec:conclusions}, we conclude the paper. 

\section{DoubleR Overview}
\label{sec:overview}

\subsection{Motivation}
\label{subsec:motivation}

Practical data centers are susceptible to both {\em independent} and
{\em correlated} node failures \cite{ford10,cidon15}: independent node
failures mean that each node fails independently due to individual events
(e.g., disk/node crashes), while correlated node failures mean that multiple
nodes fail simultaneously due to a common disastrous event (e.g., power
outages or common switch failures).  In practice, racks are treated as the
major failure domains in which correlated node failures are likely to occur.  To
deploy erasure coding in data centers, existing approaches mostly adopt
flat block placement by placing each block of a stripe in a distinct rack
\cite{ford10,huang12,sathiamoorthy13,muralidhar14,rashmi14,rashmi15}.  This
tolerates the same numbers of node failures and rack failures, and provides
the maximum fault tolerance against both independent and correlated node
failures.

Our rationale is that rack failures are much less common than node failures in
practice \cite{ford10,muralidhar14}, so it is viable to tolerate fewer rack
failures than node failures.  Thus, instead of adopting flat block placement,
we opt for hierarchical block placement and place multiple blocks in the same
rack, so as to minimize the cross-rack repair traffic at the expense of
reduced rack-level fault tolerance (note that each block is still stored in a
distinct node for the same node-level fault tolerance).  Given the constrained
bandwidth resources for cross-rack links (see \S\ref{sec:introduction}),
minimizing the cross-rack repair traffic allows fast repair, thereby
reducing the downtime of unavailable blocks (i.e., improved availability) and
the window of vulnerability (i.e., improved durability).

In fact, the use of hierarchical block placement is also found in existing
production storage systems to mitigate the cross-rack transfer overhead.
For example, HDFS \cite{shvachko10}, which is replication-based, by default
places two replicas in one rack and one replica in a different rack.
QFS \cite{ovsiannikov13}, which supports erasure coding, provides an option
called ``in-rack placement'' to place multiple blocks in the same rack.
An open issue is how to exploit the property of hierarchical block placement
to minimize the cross-rack repair traffic in erasure-coded storage, and this
is the key motivation of this work.

\subsection{DoubleR Architecture}

Given that multiple blocks are available in a rack, DoubleR exploits a
two-layer repair approach, by first repairing as much failed data as possible
within each rack, and then combining the partially repaired results from
multiple racks to repair all failed data.  To achieve bandwidth savings,
DoubleR follows {\em regenerating codes} \cite{dimakis10}, by partitioning a
block into smaller {\em subblocks} and allowing each node to compute encoded
subblocks from its stored block during a repair operation.  DoubleR takes one
step further by {\em re-encoding} the encoded subblocks from all nodes in the
same rack to achieve additional bandwidth savings across racks beyond
regenerating codes.

\begin{figure}[t]
\centering
\includegraphics[width=0.8\linewidth]{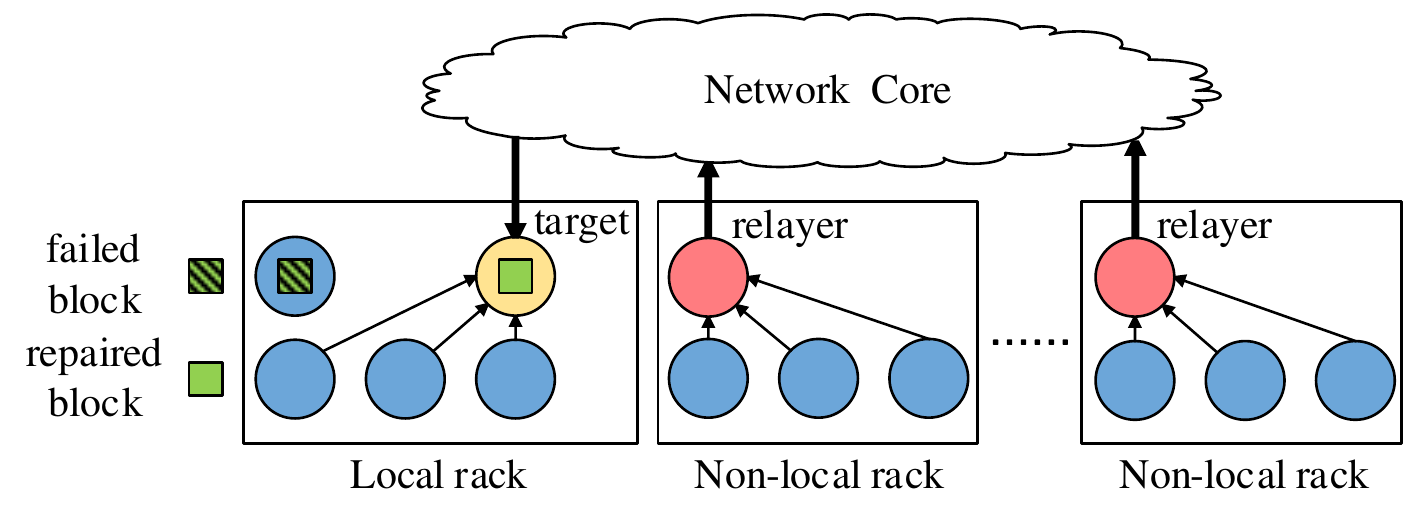}
\caption{Repair in DoubleR.}
\label{fig:overview}
\vspace{-1em}
\end{figure}

Figure~\ref{fig:overview} illustrates the repair workflow in DoubleR.
Specifically, we consider a hierarchical data center that is composed of
multiple racks, each of which contains multiple storage nodes.
Multiple nodes within the same rack are connected
by a top-of-rack switch, while multiple racks are connected by an abstraction
of switches called {\em network core} \cite{chowdhury13}.
Repairing a failed block is done by retrieving
available blocks from other nodes that reside in the same rack (called
{\em local rack}) and different racks (called {\em non-local racks}).
DoubleR selects one {\em target} node in the
local rack to be responsible for reconstructing the failed block.  It also
selects one {\em relayer} node in each non-local rack to aggregate and forward
the repaired results in that rack; typically, a relayer should be one of the
nodes that locally stores an available block for the repair so as to save
inner-rack network transfer.
In each non-local rack, each node sends encoded subblocks
to the relayer, which re-encodes the encoded subblocks.  Multiple relayers
then send the re-encoded subblocks across racks to the target, which
reconstructs the failed block.  In the deployment of DoubleR, we assign
different relayers and targets for repairing multiple failed blocks (e.g.,
when recovering all lost data of a failed node) to harness parallelism (see
\S\ref{sec:hadoop}).

DoubleR builds on the notion called {\em repair layering}, which decomposes a
repair operation into different layers (nodes and racks in our case) along the
hierarchy of a data center.  By doing so,
we can effectively mitigate the critical resource overhead,
i.e., cross-rack repair traffic.

\section{Double Regenerating Codes}
\label{sec:drc}

DoubleR builds on double regenerating codes (DRC) \cite{hu16} to realize
repair layering.  In this section, we define the notation and terminologies of
erasure coding in the context of data centers, and summarize the theoretical
findings of DRC.  We identify the connections between DRC and regenerating
codes for special cases.  We further compare DRC with existing erasure codes
for more general cases, through the numerical analysis of bandwidth savings
and the reliability modeling of mean-time-to-data-loss (MTTDL).
	
\subsection{Background}
\label{subsec:background}

We elaborate the background of erasure coding based on our discussion in
\S\ref{sec:introduction}. As multiple
stripes are independently erasure-coded, our discussion focuses on a single
stripe.  Specifically, we construct an erasure code, denoted by an $(n,k,r)$
code, with three configurable parameters $n$, $k$ (where $k<n$), and $r$
(where $r\le n$).  For each stripe, we encode $k$ original uncoded blocks of
size $B$ each into $n$ coded blocks of the same size.  For node-level fault
tolerance, we distribute the coded blocks (or blocks in short) across $n$
nodes (i.e., one block per node) that evenly reside in $r$ racks with
$n/r$ nodes each.  Here, we assume that $n/r$ is an integer.  For flat block
placement, which is used by most erasure coding deployments, we have $r = n$,
while our work addresses $r\le n$.  Unlike previous studies that typically
construct an erasure code by two parameters $n$ and $k$ only, our work
introduces the parameter $r$ to take into account rack-level fault tolerance.

We focus on erasure codes that are {\em maximum distance separable (MDS)},
meaning that any $k$ out of $n$ blocks suffice to reconstruct original
uncoded data.  MDS codes are {\em storage-optimal}, meaning that they minimize
storage redundancy (i.e., $n/k$ times the original data size).  Examples of MDS
codes include Reed-Solomon (RS) codes \cite{reed60}, which have been widely
deployed in production storage systems
\cite{hdfsraid,ford10,muralidhar14,ovsiannikov13}, as well as minimum-storage
regenerating (MSR) codes \cite{dimakis10}, which minimize the repair traffic
subject to the minimum storage redundancy.  In this paper, when we perform
comparisons with regenerating codes on repair performance, we focus on
MSR codes.  Note that some non-MDS codes are also proposed to mitigate the
repair traffic at the expense of higher storage redundancy.  Examples
include minimum-bandwidth regenerating (MBR) codes \cite{dimakis10} and
locally repairable codes \cite{huang12,sathiamoorthy13}.

In addition, we focus on {\em systematic} codes, meaning that $k$ out of $n$
coded blocks are in original uncoded form.  As opposed to non-systematic codes
(i.e., all blocks are in coded form), systematic codes allow a storage system
to directly access data without decoding.  We call the $k$ uncoded blocks as
{\em data blocks}, while the remaining $n-k$ coded blocks as
{\em parity blocks}.  For brevity, if the context is clear, we simply refer to
both data and parity blocks as ``blocks''.

\paragraph{Repair:} As in previous studies
\cite{dimakis10,khan12,huang12,rashmi14,rashmi15}, this paper focuses on
optimizing the single-failure repair, which refers to either repairing a
single unavailable block of a stripe in a degraded read operation, or
repairing all blocks of multiple stripes in a single node (i.e., one block per
stripe) in a node recovery operation.
Single-failure repair is the most common repair scenario in practice
\cite{huang12,rashmi13}.  Suppose that the target repairs a single failed
block.  In classical RS codes \cite{reed60}, the target retrieves $k$ blocks
from $k$ available nodes.  Thus, the repair traffic of RS codes per failed
block (of size $B$) is:
\begin{equation}
\begin{aligned}
B \cdot k.
\label{eqn:rs}
\end{aligned}
\end{equation}

MSR codes \cite{dimakis10} minimize the repair traffic, while achieving the
same minimum storage redundancy as RS codes (i.e., MDS).  To repair a single
failed block, each of the $n-1$ available nodes can partition a block into
$n-k$ subblocks and send an encoded subblock of size $B/(n-k)$ to the
target\footnote{MSR codes allows fewer than $n-1$ available nodes to send
encoded information for repair, at the expense of higher repair traffic.}.
The repair traffic of MSR codes per failed block \cite{dimakis10} (which is
provably minimum) is:
\begin{equation}
\begin{aligned}
B \cdot \frac{n - 1}{n - k}.
\label{eqn:min_msr}
\end{aligned}
\end{equation}

For data centers, our objective is to minimize the cross-rack repair
traffic, which is the major bottleneck in a data center (see
\S\ref{sec:introduction}).  Since each rack stores $n/r$ blocks, this work
mainly addresses the case where (1) $n/r \le k$ and (2) $n/r \le n-k$.
Case~(1) states that each rack has at most $k$ blocks, implying that
repairing a failed block must retrieve at least one available block across
racks. Case~(2) states that each rack has at most $n-k$ blocks, implying
that a single rack failure does not introduce data loss; in other words, a
data center can tolerate at least a single rack failure.  DRC \cite{hu16} is
shown to achieve the minimum cross-rack repair traffic per failed block is:
\begin{equation}
\begin{aligned}
B \cdot \frac{r - 1}{r-\lfloor kr/n \rfloor}.
\label{eqn:min_drc}
\end{aligned}
\end{equation}

If we distribute blocks across $r = n$ racks as in flat block placement,
Equation~(\ref{eqn:min_drc}) reduces to the minimum repair traffic of MSR
codes in Equation~(\ref{eqn:min_msr}).

\paragraph{Connections with regenerating codes:}
We point out that the minimum cross-rack repair traffic in
Equation~(\ref{eqn:min_drc}) can be achieved by MSR codes for specific
settings of parameters, as shown in the following theorem.

\begin{theorem}
\label{theorem:msr}
MSR codes can achieve the minimum cross-rack repair traffic for general $n$
and $k$ with $r=\frac{n}{n-k}$, assuming that $n$ is divisible by $n-k$ (i.e.,
$r$ is an integer).
\end{theorem}
\paragraph{Proof:}  We deploy MSR codes following hierarchical block placement
by setting $r = \frac{n}{n-k}$, so
each rack has $n-k$ blocks in $n-k$ different nodes.  To repair a failed
block, the target retrieves $n-k-1$ encoded subblocks from its local rack and
$(r-1)(n-k)$ encoded subblocks from non-local racks.  Since each encoded
subblock has size $\frac{B}{n-k}$, the cross-rack repair traffic of MSR codes
is $(r-1)(n-k)\cdot \frac{B}{n-k} = (r-1)B$.

Given that $r=\frac{n}{n-k}$, we can show that $k=n-\frac{n}{r}$, and hence
$\lfloor\frac{kr}{n}\rfloor = \lfloor\frac{(n-n/r)r}{n}\rfloor = r-1$.  Thus,
we can express the minimum cross-rack repair traffic in
Equation~(\ref{eqn:min_drc}) as $B \cdot \frac{r-1}{r-(r-1)} = (r-1)B$.
In other words, the cross-rack repair traffic of MSR codes is in fact the
minimum.  $\hfill\Box$

\subsection{Examples}
\label{subsec:examples}

Based on Equation~(\ref{eqn:min_drc}), we provide examples to motivate how DRC
reduces the cross-rack repair traffic over MSR codes.  Here, we fix $n=6$ and
$k=3$ in our examples.

\begin{figure}[t]
\centering
\begin{tabular}{c}
\includegraphics[width=\linewidth]{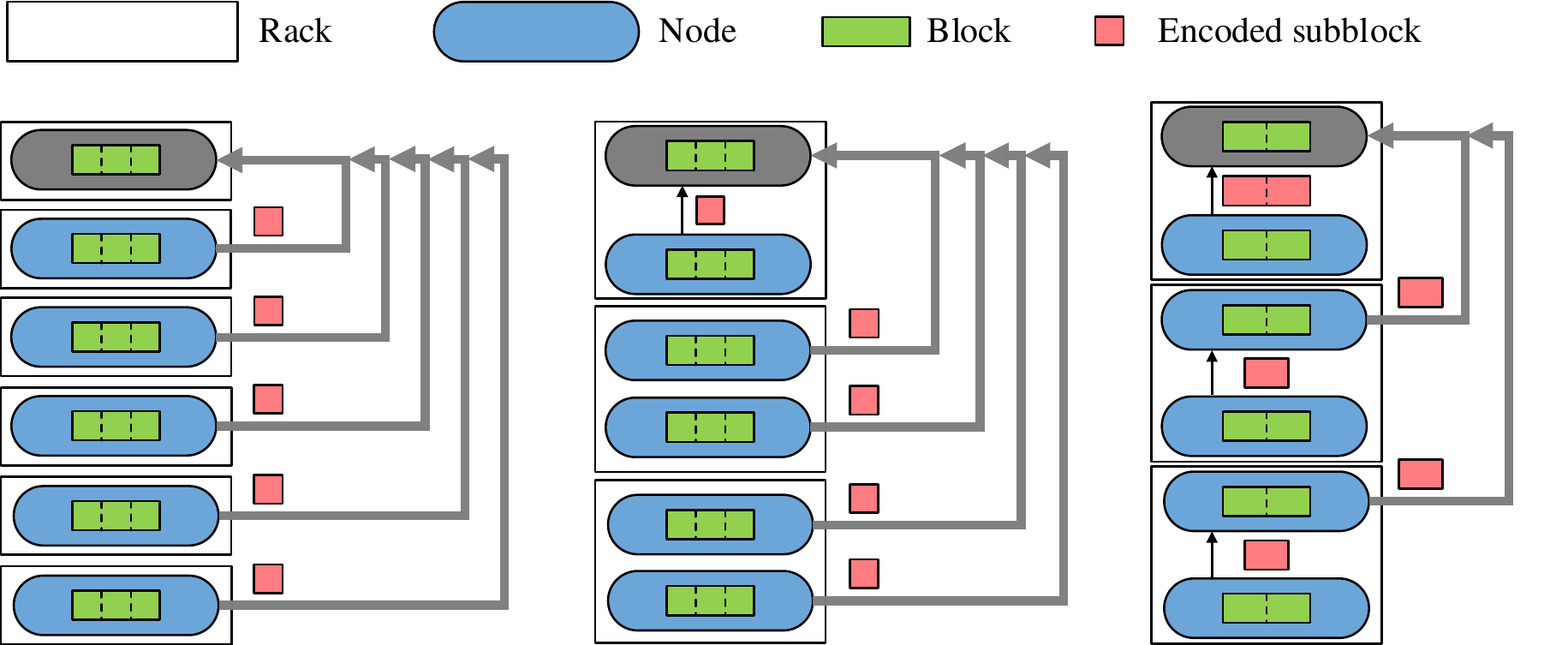}
\vspace{3pt}\\
\mbox{\small(a) MSR(6,3,6)\hspace{1in}(b) MSR(6,3,3)\hspace{1in}(c)
DRC(6,3,3)}
\end{tabular}
\vspace{-6pt}
\caption{Motivating examples.}
\label{fig:motivating}
\end{figure}

Suppose that we deploy MSR codes using flat block placement.  We
set $(n,k,r) = (6,3,6)$, and denote the code by MSR(6,3,6).
Figure~\ref{fig:motivating}(a) shows the block placement of MSR(6,3,6).  To
repair a failed block, each node in a distinct rack partitions its stored
block into $n-k=3$ subblocks and sends one encoded subblock of size
$\frac{B}{n-k} = B/3$ \cite{dimakis10}.  From Equation~(\ref{eqn:min_drc}),
the cross-rack repair traffic of MSR(6,3,6) is $5B/3$.

Clearly, we can also deploy regenerating codes using hierarchical block
placement.  We set $(n,k,r) = (6,3,3)$ by placing two blocks per rack, and
denote the code by MSR(6,3,3).  Figure~\ref{fig:motivating}(b) shows how
MSR(6,3,3) repairs a failed block.  We see that the target can retrieve
one encoded subblock from the local rack, while it still needs to retrieve
four encoded subblocks from non-local racks.  Thus, the cross-rack repair
traffic of RC(6,3,3) is reduced to $4B/3$.

DRC takes advantage of hierarchical block placement by re-encoding the encoded
subblocks in the relayer of each rack.  We again set $(n,k,r) = (6,3,3)$ and
denote the code by DRC(6,3,3).  Figure~\ref{fig:motivating}(c) shows how
DRC(6,3,3) repairs a failed block.  From Equation~(\ref{eqn:min_drc}), the
cross-rack repair traffic of DRC(6,3,3) is further reduced to $B$.

\subsection{Numerical Analysis}
\label{subsec:numerical}

We present numerical results to demonstrate the benefits of DRC in minimizing
the cross-rack repair traffic for different cases of $(n,k,r)$.  We consider
the following erasure codes:
\begin{itemize}[leftmargin=*]
\item
{\bf RS:} To repair a failed block in RS codes \cite{reed60}, the target
retrieves $k$ available blocks.  If $r < n$ (i.e., hierarchical block
placement), we assume that the target first retrieves $n/r - 1$ available
blocks from the local rack, followed by retrieving the remaining $k-(n/r-1)$
blocks from non-local racks, so as to make the cross-rack repair traffic as
low as possible for RS codes.  We use RS codes as the baseline. 
\item
{\bf MSR:}  We consider two parameter settings whose systematic MSR code
constructions have been proposed in the literature: (1) $n-k=2$
\cite{tamo13,pamies16} and (2) $n=2k$ \cite{rashmi11,shah12a,rashmi15}.  
Note that for $n-k=2$ and $r=n/2$, MSR codes achieve the same cross-rack
repair traffic as DRC (see Theorem~\ref{theorem:msr}).  
We include them for completeness. 
\item
{\bf DRC}: We consider two parameter settings: (1) general $(n,k)$ with
$r=n/(n-k)$ and (2) $(n,k,r) = (3z,2z-1,3)$ for $z\ge 2$.  In
\S\ref{sec:practical}, we provide code constructions for both settings. 
\end{itemize}

\begin{figure}[!t]
\centering
\begin{tabular}{ccc}
\includegraphics[width=1.7in]{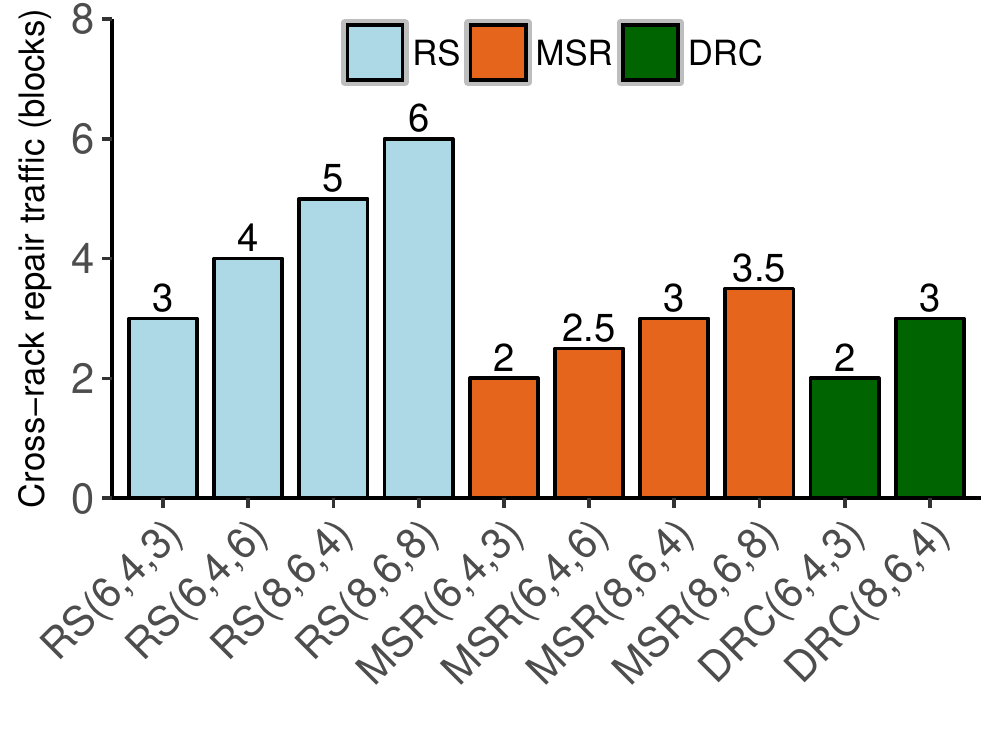} &
\includegraphics[width=1.7in]{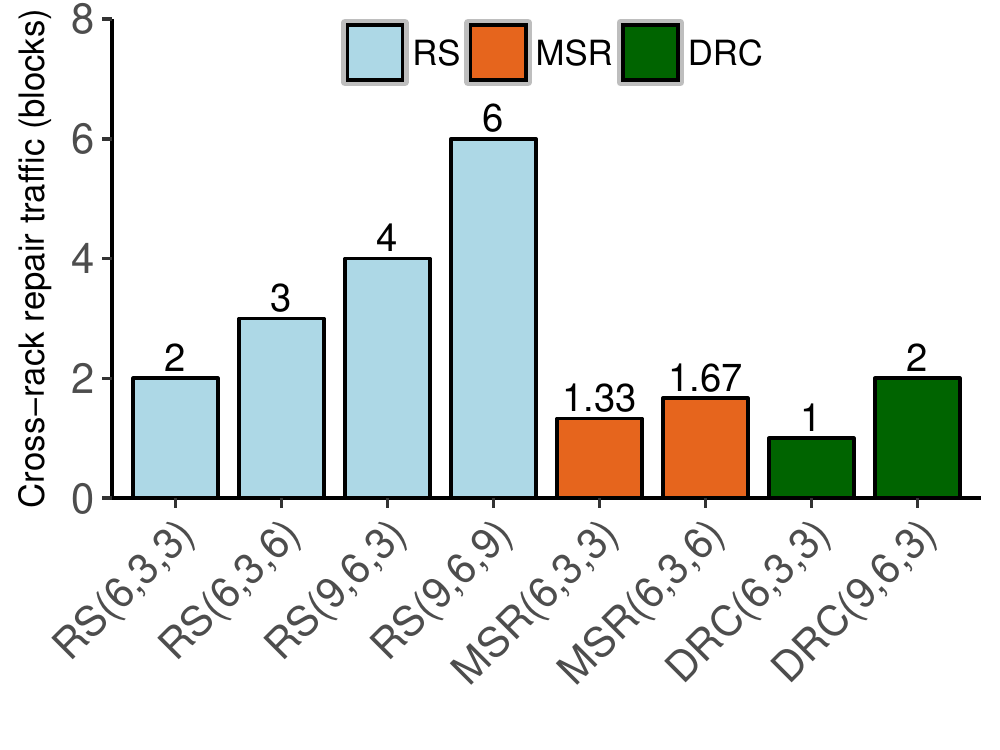} &
\includegraphics[width=1.7in]{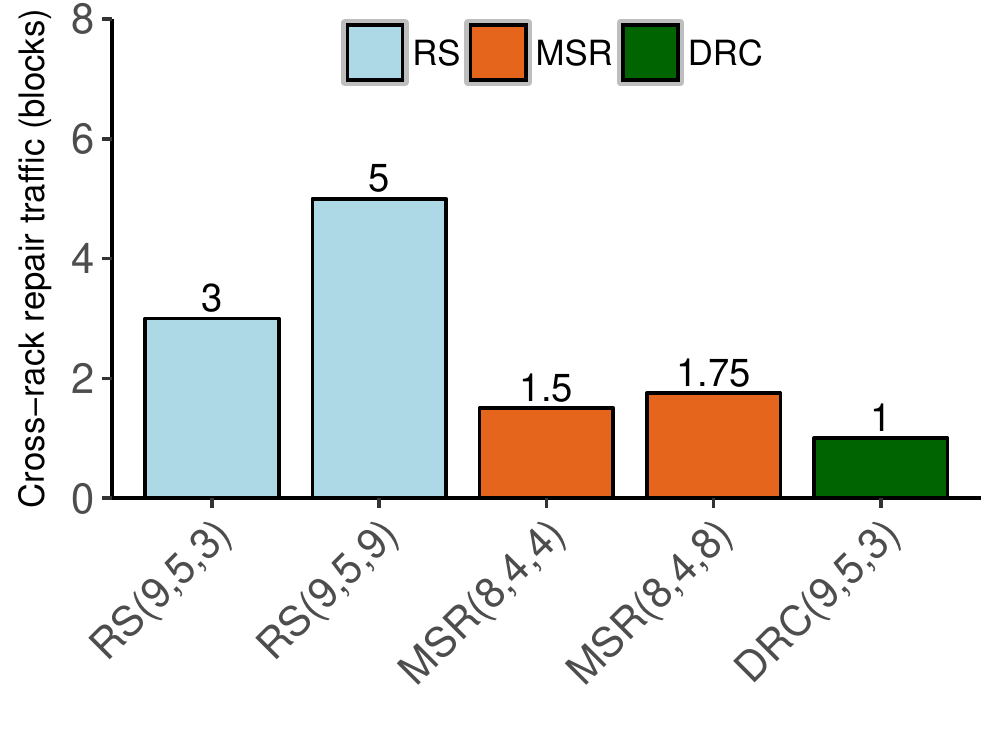} 
\vspace{-6pt}\\
\mbox{\small (a) $n-k=2$} &
\mbox{\small (b) $n-k=3$} &
\mbox{\small (c) $n-k=4$} 
\end{tabular}
\vspace{-6pt}
\caption{Numerical results of cross-rack repair traffic (in blocks) for
repairing a failed block under different erasure code configurations,
grouped by the same $n-k$.}
\label{fig:numerical}
\end{figure}

Figure~\ref{fig:numerical} shows the numerical results of cross-rack repair
traffic (in units of blocks) for repairing a failed block under
different configurations of erasure codes, which we group by the same $n-k$
(i.e., the same number of node failures that can be tolerated).  We make the
following observations:
\begin{itemize}[leftmargin=*]
\item
As expected, there is a storage-bandwidth trade-off.  For a given code with
the same $n-k$, the cross-rack repair traffic increases when the storage
redundancy (i.e., $n/k$) decreases.  For example, RS(8,6,8) has 11.3\% less
storage redundancy than RS(6,4,6), but 50\% higher cross-rack repair traffic.  
\item
For a given code and the same $n$ and $k$, hierarchical block placement
reduces the cross-rack repair traffic over flat block placement, at the
expense of reduced rack-level fault tolerance.  For example, RS(6,4,3) incurs
25\% less cross-rack repair traffic than RS(6,4,6); MSR(6,4,3) incurs 20\%
less cross-rack repair traffic than MSR(6,4,6).
\item
For the same $(n,k,r)$, DRC incurs less cross-rack repair traffic than RS
codes.  The percentage reduction increases with $n-k$.  For example,
DRC(9,5,3) incurs 66.7\% less cross-rack repair traffic than RS(9,5,3). 
\item
For $n-k\ge 3$, DRC incurs not only less cross-rack repair traffic, but also
less storage redundancy than MSR codes.  For example, DRC(9,5,3) incurs 33.3\%
less cross-rack repair traffic and 20\% less storage redundancy than
MSR(8,4,4). 
\end{itemize}

\subsection{Reliability Analysis}
\label{subsec:reliability}

Recall that DRC leverages hierarchical block placement to trade rack-level
fault tolerance for the minimum cross-rack repair traffic.  We now study the
reliability trade-off of DRC due to hierarchical block placement.  Here, we
analyze the mean-time-to-data-loss (MTTDL) metric via the standard Markov
modeling, as used by many previous studies (e.g., \cite{ford10,huang12,
sathiamoorthy13,silberstein14,cidon15}).  Although the effectiveness of
Markov-based reliability analysis is debatable \cite{greenan10}, we believe
that it suffices for providing preliminary insights on reliability for this
work.
	
\begin{figure*}[!t]
\centering
\begin{tabular}{c@{\ }c}
\includegraphics[width=2.6in]{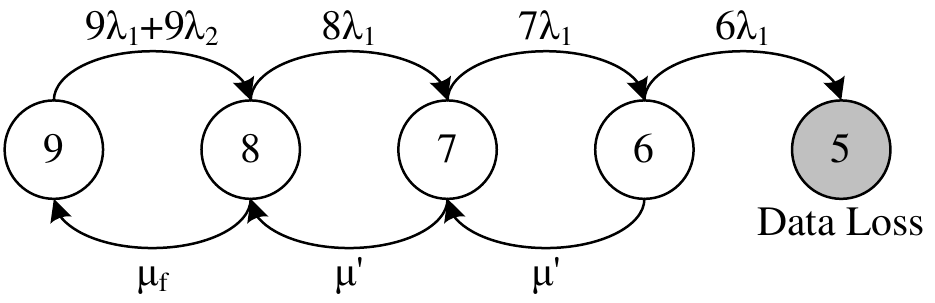} &
\includegraphics[width=2.6in]{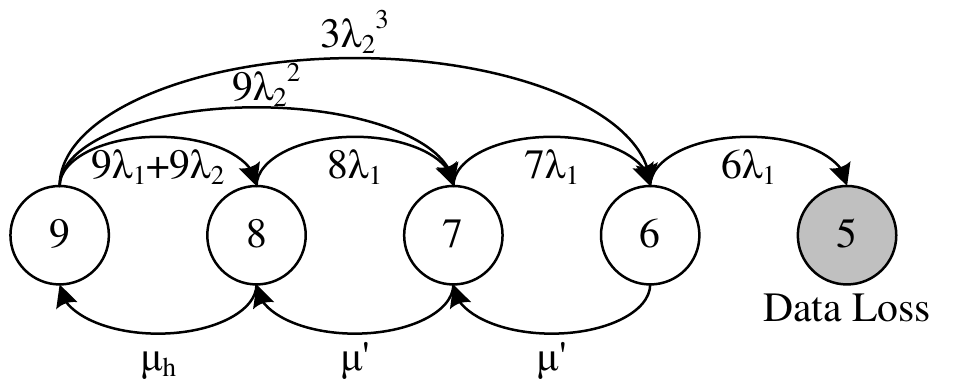} \\
\mbox{\small (a) Flat block placement $(n,k,r)=(9,6,9)$} &
\mbox{\small (b) Hierarchical block placement: $(n,k,r)=(9,6,3)$}
\end{tabular}
\vspace{-6pt}
\caption{Markov models for flat block placement and hierarchical block
placement.}
\label{fig:reliability}
\end{figure*}

\paragraph{Model:}  In our analysis, we fix $n=9$ and $k=6$
(which are also used by QFS \cite{ovsiannikov13}).
Figure~\ref{fig:reliability} shows the Markov models for
flat block placement with $(n,k,r)=(9,6,9)$ and hierarchical block placement
with $(n,k,r)=(9,6,3)$ (in \S\ref{subsec:family1}, we provide a DRC
construction for $(9,6,3)$).  Suppose that we distribute blocks of
multiple stripes across $n$ nodes. Each state represents the number of
available nodes, i.e., State~9 means that all nodes are healthy, while
State~5 implies data loss.  We assume that inter-failure and inter-repair
times are exponentially distributed.

We model both independent and correlated node failures (see
\S\ref{subsec:motivation}).  For independent node failures, let $\lambda_1$ be
the independent failure rate of each node.  The state transition rate from
State~$i$ to State~$i-1$, where $6 \leq i \leq 9$, is $i\lambda_1$, since any
of the $i$ nodes in State~$i$ fails independently.  For correlated node
failures, we consider the scenario where each rack (the largest failure domain
in our case) experiences a power outage that brings down a fixed fraction of
nodes simultaneously \cite{shvachko10,cidon15}.  In our modeling, we assume
that each node fails with a probability that is equal to the fraction of nodes
being brought down by a power outage. 
Let $\lambda_2$ be the failure rate of each node due to correlated node
failures.  We assume that correlated node failures are rare and only occur
when all nodes are healthy (i.e., State~9); in other words, a data center
operating in degraded mode is only subject to independent node failures.  This
assumption also simplifies our analysis.  Thus, for flat block placement (see
Figure~\ref{fig:reliability}(a)), the state transition rate from State~9 to
State~8 adds $9\lambda_2$, as each node residing in a distinct rack can fail
due to correlated node failures.  For hierarchical block placement (see
Figure~\ref{fig:reliability}(b)), 
there are three cases: (1) from State~9 to State~8, the
state transition rate adds $3\cdot(3\lambda_2)=9\lambda_2$, as a node failure
can occur in any one of three nodes in any one of the three racks; (2) from
State~9 to State~7, the state transition rate is
$3\cdot(3\lambda_2^2)=9\lambda_2^2$,
since a two-node failure can occur in any two of three nodes in any one of
the three racks; and (3) from State~9 to State~6, the state transition rate is
$3\lambda_2^3$ since a three-node failure can occur in any one of the three
racks.

To model repair, we assume that the repair times are proportional to the
amount of repair traffic.  When there is only one single failed node,
let $\mu_f$ and $\mu_h$ be the repair rates of a failed node from
State~8 to State~9 in flat block placement and hierarchical block placement,
respectively.  When there are multiple failed nodes, we assume
that we repair one node at a time (similar to the analysis in
\cite{huang12,sathiamoorthy13})
and the repair is done by retrieving the size of original data in both
placement schemes, and let $\mu'$ be the repair rate for each node from
State~$i$ to State~$i+1$, where $6\le i\le 8$.

We can configure the parameters as follows.  For $\lambda_1$, we assume that
the mean-time-to-failure (MTTF) of a node is in the range of a few years
\cite{schroeder07} (e.g., $1/\lambda_1=4$~years \cite{sathiamoorthy13}).
For $\lambda_2$, we follow the assumption that a power
outage occurs once a year and 0.5-1\% of nodes fail after a power outage
\cite{shvachko10}; in this case, the MTTF of a node due to a power
outage is $0.5\% \le \lambda_2 \le 1\%$ (per year).
For repair, let $\gamma$ be the
available cross-rack bandwidth, $S$ be the storage capacity of a node, and $C$
be the repair traffic per unit of repaired data.  For example, for a
single-node repair in flat block placement, $C=8/3$ if MSR codes are used
(see Equation~\ref{eqn:min_msr}), so $\mu_f = \gamma/(8S/3)$.  For a
single-node repair in hierarchical block placement, $C=2$ if DRC is used (see
Equation~\ref{eqn:min_drc}), so $\mu_h = \gamma/(2S)$.
When there are multiple failed nodes, each
failed node is repaired from the available blocks of any $k$ nodes (i.e.,
the MDS property), so $C=k=6$ and $\mu'=\gamma/(6S)$.
	
\paragraph{Analysis:}  We now evaluate the MTTDLs of both block placement
schemes.  We consider the scenarios with (1) independent node failures only
(i.e., $\lambda_2=0$) and (2) both independent and correlated node failures,
in which we set $\lambda_2=0.5\%$ (per year).  We also fix $S=1$~TiB.

We show the MTTDL results for two parameter settings: (1) we fix
$\gamma=1$~Gb/s \cite{sathiamoorthy13} and vary $1/\lambda_1$ from 2 to
10 years (see Table~\ref{tab:mttdl_lambda}), and (2) we fix
$1/\lambda_1=4$~years \cite{sathiamoorthy13} and vary $\gamma$ from 200~Mb/s
to 2~Gb/s (see Table~\ref{tab:mttdl_gamma}).  Overall, with independent node
failures only, hierarchical block placement achieves higher MTTDL than flat
block placement (by around 33\%) due to the minimized cross-rack repair
traffic in repairing a single node failure. However, when correlated node
failures are included, the MTTDL drop in hierarchical block placement is much
more obvious than in flat block placement.

Specifically, in the presence of correlated node failures, hierarchical block
placement has a relatively higher MTTDL than flat block placement when
independent node failures are more frequent, in which case the repair rate
plays a more dominant role in MTTDL.  For example, in
Table~\ref{tab:mttdl_lambda}, when $1/\lambda_1=2$~years
and there are correlated node failures, the MTTDL of flat block
placement is $2.54\times10^6$~years, while that of hierarchical block
placement is $3.28\times10^6$~years (29\% higher).  On the other hand,
hierarchical block placement has less MTTDL than flat block placement when
$1/\lambda_1$ increases or $\gamma$ increases (e.g., $1/\lambda_1\ge 6$~years
in Table~\ref{tab:mttdl_lambda} and $\gamma=2$~Gb/s in
Table~\ref{tab:mttdl_gamma}).  In this case, the improvement
due to the minimized cross-rack repair traffic becomes less important.
Nevertheless, the overall impact of failures is also low and hierarchical
block placement already achieves a fairly high MTTDL (e.g., over $10^8$ years
for $1/\lambda_1\ge 6$~years as shown in Table~\ref{tab:mttdl_lambda}). 

\begin{table}[!t]
\centering
\caption{MTTDLs of flat block placement and hierarchical block placement for
different values of $1/\lambda_1$ (years) and $\gamma_1=1$~Gb/s.}
\label{tab:mttdl_lambda}
\begin{tabular}{|l|c|c|c|c|c|}
\hline
$1/\lambda_1$ (years)   &  2     & 4    & 6       & 8      & 10   \\ \hline
Flat, w/o correlated & 2.56E+06 & 4.08E+07 & 2.06E+08 & 6.52E+08 & 1.59E+09 \\ \hline
Flat, w/ correlated  & 2.54E+06 & 4.00E+07 & 2.00E+08 & 6.27E+08 & 1.51E+09 \\ \hline
Hierarchical, w/o correlated & 3.41E+06 & 5.44E+07 & 2.75E+08 & 8.69E+08 & 2.12E+09 \\ \hline
Hierarchical, w/ correlated  & 3.28E+06 & 4.69E+07 & 1.96E+08 & 4.81E+08 & 8.80E+08 \\ \hline
\end{tabular}
\end{table}
\begin{table}[!t]
\centering
\caption{MTTDLs of flat block placement and hierarchical block placement for
different values of $\gamma$ (Gb/s) and $1/\lambda_1=4$ years.}
\label{tab:mttdl_gamma}
\begin{tabular}{|l|l|l|l|l|}
\hline
$\gamma$ (Gb/s)      & 0.2      & 0.5      & 1        & 2        \\ \hline
Flat, w/o correlated & 3.32E+05 & 5.12E+06 & 4.08E+07 & 3.26E+08 \\ \hline
Flat, w/ correlated & 3.26E+05 & 5.02E+06 & 4.00E+07 & 3.19E+08 \\ \hline
Hierarchical, w/o correlated & 4.42E+05 & 6.82E+06 & 5.44E+07 & 4.34E+08 \\ \hline
Hierarchical, w/ correlated & 4.25E+05 & 6.33E+06 & 4.69E+07 & 3.09E+08 \\ \hline
\end{tabular}
\end{table}

\section{Practical DRC Constructions}
\label{sec:practical}

It is shown in \cite{hu16} that DRC can be constructed via random linear
codes. A major drawback is that such a construction is not practical, as it is
non-systematic (i.e., all blocks are in coded form).  This implies that extra
decoding is needed to access any coded block.  In this section, we provide
practical DRC constructions that are suitable for real deployment.

\subsection{Goals}
\label{subsec:goals}

Our practical DRC constructions aim for several design goals:
\begin{enumerate}[leftmargin=*] \itemsep=0pt \parskip=0pt
\item
{\bf Theoretical guarantees:} Each construction is MDS (and hence
storage-optimal) and minimizes the cross-rack repair traffic.
\item
{\bf Systematic:}  The original data blocks are kept after encoding.
\item
{\bf Exact-repair:}  Each reconstructed block has the same content as the
original failed block.
\item
{\bf Small finite fields:}  The arithmetic operations of encoding are done
over the Galois field GF($2^8$); in other words, the encoding can be done in
units of bytes \cite{greenan08}.
\item
{\bf Small redundancy:}  Each construction can achieve storage redundancy
below 2$\times$.
\item
{\bf Polynomial number of subblocks per block:}  The number of subblocks per
block is polynomial with respect to the number of original data blocks of a
stripe (i.e., $k$); this reduces the access overhead to subblocks.
\item
{\bf Reduced inner-rack repair traffic:}  The amount of traffic that a
relayer receives from all available nodes within the same rack is no more than
that it sends out to the target across the racks.
\item
{\bf Balanced cross-rack repair traffic:} Each relayer sends the same amount
of cross-rack traffic during repair.
\end{enumerate}

The design goals have the following implications.
Goal~1 ensures that our practical DRC constructions preserve the theoretical
guarantees as proven in \cite{hu16}.
Goals~2-4 improve non-systematic regenerating codes \cite{dimakis10} and DRC
\cite{hu16}, both of which require that the Galois field size needs to be
sufficiently large to provide theoretical guarantees.
Goal~5 improves existing systematic regenerating codes including MISER codes
\cite{shah12a}, Product-Matrix (PM) codes \cite{rashmi11}, and PM-RBT codes
\cite{rashmi15}, all of which require the redundancy be at least 2$\times$.
Goal~6 improves Butterfly codes \cite{pamies16}, which have $2^{k-1}$ subblocks
per block (i.e., exponential with $k$).
Goal~7 ensures that by limiting the inner-rack repair traffic, the
cross-rack repair traffic is the most dominant factor in the repair
performance.
Finally, Goal~8 ensures that the repair operation is load-balanced across all
racks.

To this end, we propose two families of practical DRC constructions for
different possible configurations of $(n,k,r)$.  Both families can tolerate a
variable number of node failures, while tolerating only a single-rack failure
to trade for the minimum cross-rack repair traffic.
Family~1, called DRC$(n,k,\frac{n}{n-k})$, supports general $n$ and $k$ ($k <
n$) with $r=\frac{n}{n-k}$, provided that $r$ is an integer.  It can be
configured with low storage redundancy (e.g., 1.33$\times$ for $(8,6,4)$).
Family~2, called DRC$(3z,2z-1,3)$, supports any integer $z\!\ge\!2$.  Its
redundancy is 1.5-2$\times$ that is generally higher than that of Family~1,
but achieves less cross-rack repair traffic.  Our current DoubleR prototype
has implemented DRC(6,4,3), DRC(8,6,4), and DRC(9,6,3) for Family~1, and
DRC(6,3,3) and DRC(9,5,3) for Family~2.

One key property of both families is that their encoding/decoding operations
are based on the classical RS codes \cite{reed60}, which have been well
studied and widely deployed in production
\cite{hdfsraid,ford10,muralidhar14,ovsiannikov13}.  Thus, we can exploit the
theoretical guarantees provided by RS codes (e.g., the MDS property).  A key
challenge is how to augment RS codes to satisfy the design goals listed in
\S\ref{subsec:goals}, which we address below.

Before we present the two families of DRC constructions, we remark that a
number of recent studies \cite{ye17a,ye17b,goparaju17,sasidharan16}
have proposed MSR code constructions that support general $n$ and $k$ and
also achieve Goals~1-6 stated above.  Recall that Family~1 has the same
parameters in Theorem~\ref{theorem:msr}.  Thus, we can directly use the
recently proposed MSR code constructions to minimize the cross-rack repair
traffic.  Nevertheless, such MSR code constructions are mainly studied from a
theoretical standpoint, and their implementations and evaluations are still
open issues.  In contrast, Family~1 can be realized by RS codes, which are well
known in practice.

\subsection{Family 1: DRC$(n,k,\frac{n}{n-k})$}
\label{subsec:family1}

\begin{figure*}[!t]
\centering
\includegraphics[width=.95\linewidth]{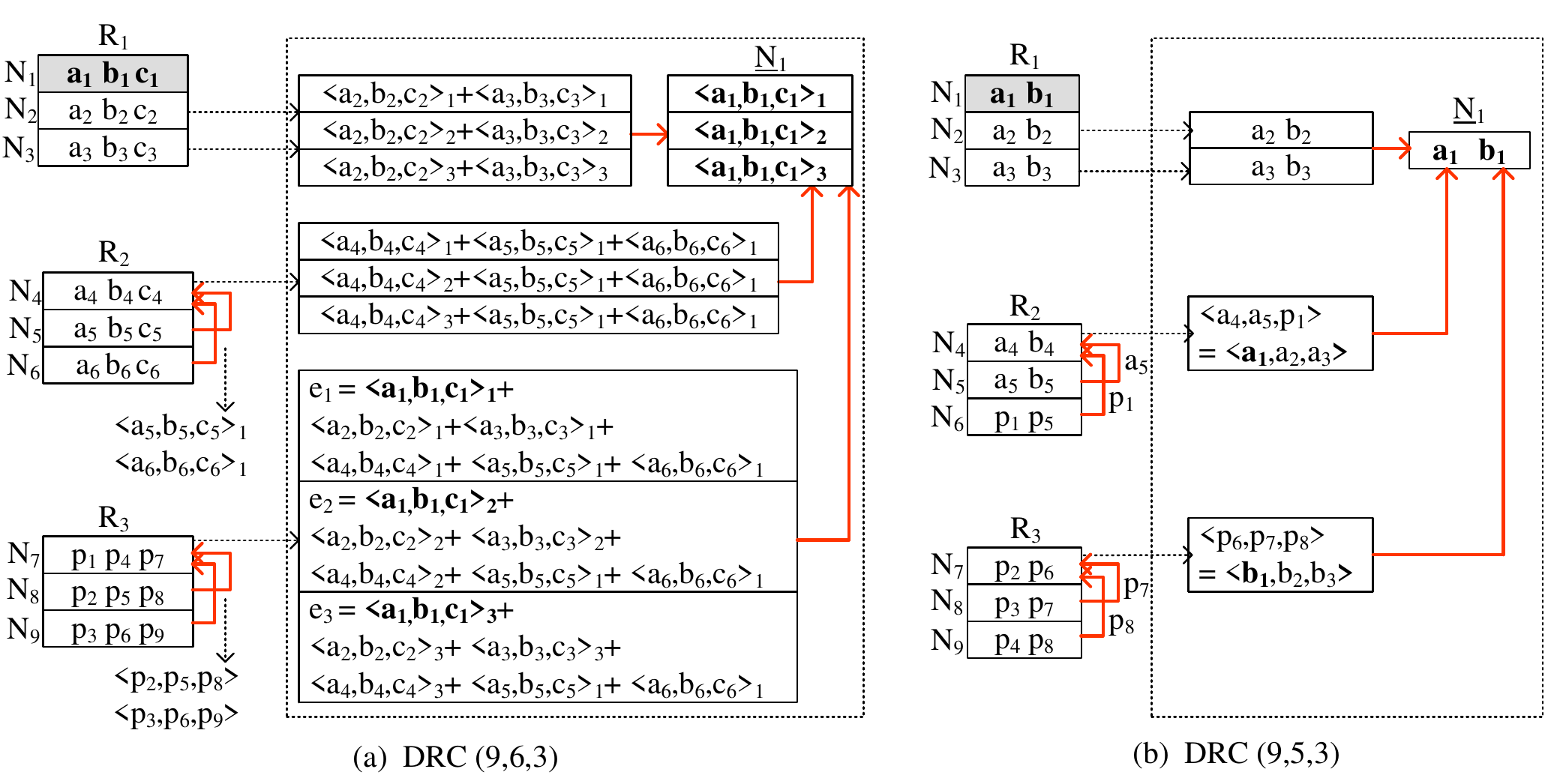}
\vspace{-.5em}
\caption{Two families of DRC. The red (solid) arrows refer to the network
transfers of encoded subblocks.}
\label{fig:963953}
\vspace{-1em}
\end{figure*}

Family~1 integrates RS codes with {\em interference alignment} \cite{shah12a},
which makes the reduction of inner-rack repair traffic (Goal~7) and the
balance of cross-rack repair traffic across multiple racks (Goal~8)
possible.  We use DRC(9,6,3) as an example, as shown in
Figure~\ref{fig:963953}(a).

\paragraph{Construction:}  For each stripe, we collect a set of $k$ data
blocks, and divide each block into $n-k$ subblocks, called {\em data subblocks}.
We group the subblocks at the same offset of all data blocks into a set, so
there are $n-k$ sets in total.  We construct $n-k$ parity blocks from the $k$
data blocks by encoding each set of $k$ data subblocks into $n-k$ coded
subblocks, called {\em parity subblocks}, using RS encoding.

For example, consider DRC(9,6,3) in Figure~\ref{fig:963953}(a).  We have nine
nodes, denoted by $N_1$, $N_2$, $\cdots$, $N_9$, that are placed across three
racks, denoted by $R_1$, $R_2$, and $R_3$.  We divide $k=6$ data blocks into
three sets of subblocks \{$a_1,\dots,a_6$\}, \{$b_1,\dots,b_6$\}, and
\{$c_1,\dots,c_6$\}, which are stored in the first six nodes $N_1$ to $N_6$.
We perform RS encoding to encode each of the three sets of data subblocks and
respectively form three sets of parity subblocks, denoted by $\{p_1,p_2,p_3\}$,
$\{p_4,p_5,p_6\}$, and $\{p_7,p_8,p_9\}$.
The parity blocks are stored in the remaining three nodes $N_7$ to $N_9$.
Since the original $k$ data blocks can be reconstructed from any $k$ blocks of
a stripe via RS decoding, the MDS property is preserved.

\paragraph{Repair idea:}  We first describe the main idea of repairing a data
block, and discuss how it also applies to repairing a parity block.

Without loss of generality, we repair a data block in $N_1$. Each
relayer (say, $N_4$ and $N_7$) sends the target, denoted by
$\underline{N}_1$, three encoded subblocks of size $B/3$ each.  Based
on DoubleR, the target $\underline{N}_1$ can obtain data from $N_2$ and $N_3$
of the local rack and relayers $N_4$ and $N_7$ of the non-local racks to
repair the lost subblocks \{$a_1,b_1,c_1$\}.  Obviously, $N_2$, $N_3$, and
$N_4$ cannot provide any information pertaining to $\{a_1,b_1,c_1\}$, while
only $N_7$ can provide three encoded subblocks, denoted by $\{e_1,e_2,e_3\}$,
in the form of linear combinations of all data subblocks. Thus,
$\underline{N}_1$ must be able to solve for $\{a_1,b_1,c_1\}$ from
$\{e_1,e_2,e_3\}$ by subtracting out the ``interference'' information
pertaining to the non-failed data subblocks.  We borrow the idea of
interference alignment \cite{shah12a} to make the interference information
formed by a number of {\em aligned} linear combinations, which are either the
same or scalar multiples of each other.  Instead of solving for individual
subblocks the repair operation now cancels out aligned linear combinations
through linear algebra.

To repair a parity block (e.g., in $N_7$), we represent its parity subblocks
$p_1$, $p_4$, $p_7$ by $a'_1$, $b'_1$, and $c'_1$, respectively.  Due to RS
encoding, the data subblocks $a_1$, $b_1$, and $c_1$ are in fact the linear
combinations of $\{a'_1,a_2,\dots,a_6\}$, $\{b'_1,b_2,\dots,b_6\}$, and
$\{c'_1,c_2,\dots,c_6\}$, respectively.  Thus, we can view $\{a'_1, b'_1,
c'_1\}$ as data subblocks and $\{a_1, b_1, c_1\}$ as parity subblocks.  In this
way, we can apply the same approach of repairing a data block into repairing a
parity block.

\paragraph{Repair details:} We specify the detailed steps of repairing $N_1$,
while the same methodology applies to other nodes.

(1) The relayer $N_7$ sends encoded subblocks $\{e_1,e_2,e_3\}$ to the target
$\underline{N}_1$, such that each encoded subblock comprises the same aligned
linear combination of $\{a_5,b_5,c_5,a_6,b_6,c_6\}$.  We elaborate how to form
$\{e_1,e_2,e_3\}$ below.

(1.1) $e_1$ is simply given by $p_1 + p_4 + p_7$, which can be viewed as a
linear combination of all data subblocks. Let
$e_1 = \sum_{i=1}^{6} \langle a_i,b_i,c_i \rangle_1$, where
$\langle x_1,\dots,x_m \rangle_j$ denotes the $j^{th}$ linear combination of
subblocks $\{x_1,\dots,x_m$\}.

(1.2) $e_2$ is a linear combination of
$\{p_1,p_4,p_7,\langle p_2,p_5,p_8\rangle\}$,
where $\langle p_2,p_5,p_8 \rangle$ is a linear combination sent from $N_8$ to
$N_7$.  We ensure that $e_2$ also contains
$\sum_{i=5}^{6} \langle a_i,b_i,c_i\rangle_1$,
which aligns with part of $e_1$, and is represented as
$e_2=\sum_{i=1}^{4} \langle a_i,b_i,c_i \rangle_2 + \sum_{i=5}^{6} \langle
a_i,b_i,c_i \rangle_1$.  This can be accomplished by tuning the coding
coefficients as follows.  Let $e_2 = \gamma_1 p_1 + \gamma_2 p_4 + \gamma_3
p_7 + (\gamma_4 p_2 + \gamma_5 p_5 + \gamma_6 p_8)$, where $\gamma_i$'s are
some coding coefficients.  Then we can tune $\gamma_1$ and $\gamma_4$ in such
a way that $\gamma_1 p_1 + \gamma_4 p_2$ have the same terms for $a_5$ and
$a_6$ as in $\sum_{i=5}^{6} \langle a_i,b_i,c_i\rangle_1$.  Similarly, we can
tune $\gamma_2$ and $\gamma_5$ to have the same terms for $b_5$ and $b_6$, and
tune $\gamma_3$ and $\gamma_6$ to have the same terms for $c_5$ and $c_7$.

(1.3) $e_3$ is a linear combination of
$\{p_1,p_4,p_7,\langle p_3,p_6,p_9\rangle\}$,
where $\langle p_3,p_6,p_9 \rangle$ is a linear combination sent from $N_9$ to
$N_7$.  We also ensure that $e_3$ also contains $\sum_{i=5}^{6}\langle
a_i,b_i,c_i \rangle_1$, which again aligns with part of $e_1$, by setting
$e_3=\sum_{i=1}^{4}\langle a_i,b_i,c_i \rangle_3 + \sum_{i=5}^{6} \langle
a_i,b_i,c_i \rangle_1$.  This can be done by tuning coding coefficients as
above.

(2) The relayer $N_4$ computes $\langle a_4,b_4,c_4 \rangle_1$ and retrieves
the linear combinations $\langle a_5,b_5,c_5 \rangle_1$ and
$\langle a_6,b_6,c_6 \rangle_1$ from $N_5$ and $N_6$, respectively.  It sends
$\underline{N}_1$ three encoded subblocks, each of which aligns with part of 
$e_1$, $e_2$, or $e_3$.  For example, the linear combination 
$\sum_{i=4}^{6}\langle a_i,b_i,c_i\rangle_1$ aligns with part of $e_1$. 

(3) Each of the nodes $N_2$ and $N_3$ in the local rack sends
$\underline{N}_1$ three encoded subblocks, each of which aligns with part of 
$e_1$, $e_2$, or $e_3$. 

(4) $\underline{N}_1$ cancels out the aligned linear combinations.
It now has $\langle a_1,b_1,c_1 \rangle_1$, $\langle a_1,b_1,c_1 \rangle_2$,
and $\langle a_1,b_1,c_1 \rangle_3$, which can be used to solve for
$\{a_1,b_1,c_1\}$.

\subsection{Family 2: DRC$(3z,2z\!-\!1,3)$}
\label{subsec:family2}

Family~2 differs from Family~1 by allowing a node in a non-local rack to
merely read a subblock from its local storage and send it to the relayer,
without performing encoding operations.  It follows the spirit of
repair-by-transfer \cite{shah12b} and helps reduce disk I/Os.  We use
DRC(9,5,3) as an example, as shown in Figure~\ref{fig:963953}(b).

\paragraph{Construction:} For each stripe, we collect $k$ data blocks and
divide each block into two subblocks (i.e., $2(2z-1)$ data subblocks in total).
We group the subblocks at the same offset of all data blocks into a set (i.e.,
there are two sets in total).  Each set of $2z-1$ data subblocks is
independently encoded using RS codes to generate $z+1$ parity subblocks.  We
distribute the $n=3z$ blocks across three racks, each of which contains $z$
blocks (i.e., $3z$ subblocks in total).  Like Family~1, since each set of
subblocks is encoded with RS codes, the MDS property is preserved.

For example, consider DRC(9,5,3) in Figure~\ref{fig:963953}(b) (i.e., $z=3$).
We have nine nodes $N_1, N_2, \cdots, N_9$ that are placed across three racks
$R_1$, $R_2$, and $R_3$.  The data blocks have two sets of data subblocks
$\{a_1,\cdots,a_5\}$ and $\{b_1,\cdots,b_5\}$, and we place the data blocks
in $N_1$ to $N_5$.  We encode them using RS codes to generate two sets of
parity subblocks $\{p_1,p_2,p_3,p_4\}$ and $\{p_5,p_6,p_7,p_8\}$, respectively,
and place the parity blocks in $N_6$ to $N_9$.

\paragraph{Repair idea:} As in Family~1, we only need to consider how to
repair a data block (e.g., in $N_1$), while we apply the same methodology to
repair a parity block.  Our observation is that each failed subblock can be
reconstructed by $2z-1$ subblocks from two racks only, i.e., the $z-1$ subblocks
of the same set in the local rack and the $z$ subblocks of the same set in one
of the non-local racks.  For example, in Figure~\ref{fig:963953}(b), the
failed subblock $a_1$ can be reconstructed from $\{a_2,a_3,a_4,a_5,p_1\}$,
which reside in $R_1$ and $R_2$, while the failed subblock $b_1$ can be
reconstructed from
$\{b_2,b_3,p_6,p_7,p_8\}$, which reside in $R_1$ and $R_3$.  The two relayers
(say, $N_4$ and $N_7$) only need to send information that is needed for
reconstructing $a_1$ and $b_1$, respectively, and the cross-rack repair
traffic can be shown to be minimum.  Also, note that each of $N_4, \cdots
N_9$ only needs to read a subblock from its local storage, where the subblock
size is only half of the block size.  This reduces disk I/Os.

\paragraph{Repair details:} We specify the detailed steps of repairing the
data block in $N_1$.

(1) The relayer $N_4$ in $R_2$ collects the subblocks $a_4$, $a_5$, and $p_1$
within the same rack.  It computes a linear combination of the collected
subblocks as an encoded subblock, such that both $a_4$ and $a_5$ can be canceled
out; in other words, it computes the encoded subblock as
$\langle a_4,a_5,p_1\rangle = \langle a_1,a_2,a_3\rangle$.  This can be
accomplished by simply subtracting out $a_4$ and $a_5$ from $p_1$, as $p_1$ is
a linear combination of $\{a_1,a_2,\cdots,a_5\}$.  It sends the encoded
subblock to the target $\underline{N}_1$.

(2) The relayer $N_7$ in $R_3$ collects the subblocks $p_6$, $p_7$, and $p_8$
within the same rack.  Similar to above, $N_7$ computes a linear combination
of $p_6$, $p_7$, and $p_8$ as an encoded subblock, such that both $b_4$ and
$b_5$ can be canceled out.  Thus, the encoded subblock is computed as
$\langle p_6,p_7,p_8 \rangle = \langle b_1,b_2,b_3 \rangle$, and sent to the
target $\underline{N}_1$.

(3) Both nodes $N_2$ and $N_3$ send their stored subblocks to the target
$\underline{N}_1$.

(4) $\underline{N}_1$ solves for $a_1$ and $b_1$ by canceling out $a_2$,
$b_2$, $a_3$, and $b_3$ from $\langle a_1,a_2,a_3 \rangle$ and $\langle
b_1,b_2,b_3 \rangle$ through linear algebra.

\section{Implementation}
\label{sec:hadoop}

We implement DoubleR on Facebook's HDFS \cite{hadoop20}, which integrates
HDFS-RAID \cite{hdfsraid} to support erasure coding atop Hadoop Distributed
File System (HDFS) \cite{shvachko10}.  We provide an overview of how HDFS
realizes erasure coding, and then describe how we implement DoubleR atop HDFS. 

\subsection{HDFS Overview}

HDFS organizes data as fixed-size data blocks, each of which is a basic unit
of read/write operations and has a large size (e.g., 64~MiB) to mitigate
random access overhead.  It comprises a single {\em NameNode} for managing
file operations and multiple {\em DataNodes} for storing data.

HDFS-RAID adds a {\em RaidNode} to HDFS for managing erasure-coded blocks.
The RaidNode first stores data blocks with replication, such that each data
block has multiple copies stored in distinct nodes.  It later transforms the
blocks into erasure-coded blocks in the background.  The RaidNode coordinates
the transformation via MapReduce \cite{dean04}.  Specifically, to construct an
erasure code with parameters $n$ and $k$ for a stripe, a map task of a
MapReduce job collects $k$ data blocks from different DataNodes, encodes them
into $n-k$ parity blocks, and distributes the $n$ blocks across $n$ different
DataNodes.  In addition, the RaidNode periodically checks for any failed
blocks and triggers the repair operation if needed.

\subsection{DoubleR Details}
\label{subsec:details}

We explain how we extend the Facebook's HDFS to include DoubleR, and provide
justifications for our design choices.

\paragraph{Erasure codes:}  We implemented different erasure codes based on
the parameters shown in \S\ref{subsec:numerical}, including RS codes, MSR
codes, and DRC.  For MSR codes, we implemented Butterfly codes \cite{pamies16}
for $n-k=2$ and MISER codes \cite{shah12a} for $n=2k$; both codes are
systematic codes.  For DRC, we implemented the two families of practical DRC
constructions for different combinations of $(n,k,r)$ (see \S\ref{sec:drc}).

Each erasure code is implemented in C++ using Intel's ISA-L \cite{isal}.  We
mainly use two ISA-L APIs: \texttt{ec\_init\_tables}, which specifies the
coding coefficients, and \texttt{ec\_encode\_data}, which specifies the
encoding/decoding operations.  Both APIs automatically optimize the
computations based on the hardware configurations (e.g., Intel SSE
instructions are used if supported).  We link each erasure code implementation
with Hadoop via Java Native Interface (JNI).

\paragraph{Strip size:}  In the original HDFS-RAID, a block is partitioned
into multiple {\em strips} for erasure coding, such that the strips at the
same block offsets are encoded together to form a smaller-size stripe.  Our
DoubleR implementation exploits this feature and further uses multi-threading
(see details below) to parallelize the encoding/decoding of blocks that now
span multiple smaller-size stripes.  For both regenerating codes and DRC,
each strip is divided into {\em substrips}, so that the available nodes can
send encoded substrips for repair; in other words, an encoded subblock in
regenerating codes and DRC is composed of multiple encoded substrips of a
block.  Note that if the strip size is too small, there will be heavy I/O
access overhead.  We study the impact of the strip size through experiments
(see \S\ref{sec:experiments}). 

\paragraph{Block placement:}  DoubleR groups multiple blocks belonging to the
same stripe in the same rack.  We modify the RaidNode to specify how blocks of
each stripe are stored based on the parameters $(n,k,r)$.

\paragraph{Repair operations:} DoubleR focuses on repairing single failures.
It currently supports two types of repair operations: {\em node recovery} and 
{\em degraded reads}.  A node recovery operation repairs multiple failed
blocks of a single failed node, in which each failed block belongs to a
different stripe.  We modify the RaidNode to call DoubleR for node recovery
when it detects a failed node.  On the other hand, a degraded read operation
repairs a single unavailable block.  We modify the file system client to call
DoubleR to perform a degraded read when it fails to access a block and
triggers a block missing exception. 

\paragraph{Parallelization:}  DoubleR does not leverage MapReduce for repair
as in the original HDFS-RAID; instead, its implementation embodies extensive
parallelization to speed up a repair operation and move the bottleneck to
cross-rack transfer.  First, we use multi-threading at the node level to
parallelize disk I/O, encoding/decoding, and network transfer operations.  
We also spawn multiple threads to repair multiple strips of a failed block in
parallel.  In addition, for node recovery, which involves the repair of
multiple failed blocks of a single node, we assign different relayers and
targets for different stripes to harness parallelism in a data center. 

\paragraph{Exported APIs:}  DoubleR exports three primitive APIs for a repair
operation: (1) {\tt NodeEncode}, in which a storage node computes encoded
subblocks from its locally stored block, (2) {\tt RelayerEncode}, in which a
relayer computes re-encoded subblocks from the encoded subblocks of the
storage nodes in the same rack, and (3) {\tt Decode}, in which a target
reconstructs a failed block, using the blocks from the nodes in the same rack
and blocks from the relayers in different racks.  For regenerating codes
\cite{dimakis10} and their variants \cite{shah12a,pamies16}, we only need to
implement {\tt NodeEncode} and {\tt Decode}; for DRC, we implement all three
APIs.

\section{Experiments}
\label{sec:experiments}

We present evaluation results on DoubleR from testbed experiments.  We address
the following questions:
(1) Can DRC achieve the theoretical performance (i.e., the numerical results
in \S\ref{subsec:numerical}) in a real networked environment?
(2) Does minimizing cross-rack repair traffic play a key role in
improving the overall repair performance?

\subsection{Methodology}
\label{subsec:methodology}

\paragraph{Testbed setup:} Our testbed experiments are conducted on a cluster
of 11 machines.  Each machine has a quad-core 3.4~GHz Intel Core i5-3470,
16~GiB RAM, and a Seagate ST1000DM003 7200~RPM 1~TiB SATA hard disk. All
machines are interconnected via a 10~Gb/s Ethernet switch.  We deploy
Facebook's HDFS \cite{hadoop20} on 10 machines. One machine runs both the
NameNode and RaidNode, and each of other $n$ machines runs a DataNode for an
$(n,k,r)$ code, where $n$ is up to 9 in our evaluation.

To mimic a hierarchical data center, we assign one dedicated machine called
the {\em gateway} to mimic the network core in Figure~\ref{fig:overview}.
Specifically, we partition the $n$ DataNodes into $r$ logical racks.  If one
machine in a logical rack wants to send data to another machine in a different
logical rack, its cross-rack traffic will first be redirected to the
gateway, which then relays the traffic to the destination machine; otherwise,
its inner-rack traffic will be sent directly to the destination machine
through the 10~Gb/s Ethernet.  We configure the routing table of each machine
using the Linux command {\tt route} for the traffic redirection.  In addition,
we limit the outgoing bandwidth of the gateway (i.e., the available cross-rack
bandwidth) using the Linux traffic control command {\tt tc}, so as to mimic the
oversubscription scenario (see \S\ref{sec:introduction}).

\paragraph{Default parameters:}  We study different erasure codes that we
implemented (see \S\ref{subsec:details}).
By default, we configure the block size as
64~MiB (which is also the default in Facebook's HDFS) and the strip size as
256~KiB.  One subtlety is that both MISER(6,3,3) and DRC(9,6,3) need to
partition a block (strip) into three subblocks (substrips) for repair.  To
allow even partitioning, for both cases, we configure the block size as 63~MiB
and the strip size as 252~KiB as their defaults.  We also set the default
gateway bandwidth as 1~Gb/s to simulate the available cross-rack bandwidth for
repair in production data centers \cite{sathiamoorthy13}.  We vary one of the
parameters in each of our experiments.  Our results are averaged over five
runs; we omit the variances of the results as they are insignificant based on
our evaluation.

\subsection{Microbenchmarks}
\label{subsec:microbenchmarks}

Before we measure the node recovery and degraded read performance, we first
show via microbenchmark evaluation that cross-rack transfer is indeed the most
dominant factor in the overall repair performance.
We study DRC(9,6,3) and DRC(9,5,3) as the representatives for Family~1 and
Family~2, respectively, using the default parameters.
We provide a breakdown of the repair time for a
single failed block; note that the default block sizes for DRC(9,6,3) and
DRC(9,5,3) are 63~MiB and 64~MiB, respectively.  We decompose a repair
operation into different steps, including sending data over the network and
performing local computations in different APIs (see \S\ref{subsec:details}).
We derive the expected running time of each step as follows.
\begin{itemize}[leftmargin=*]
\item
{\bf Disk read:} For both DRC(9,6,3) and DRC(9,5,3), each available node first
reads a block from its local disk.
Our measurement indicates that the disk read throughput of our testbed is
around 177~MiB/s.  Thus, the disk read times for a single block for DRC(9,6,3)
and DRC(9,5,3) are 0.354s and 0.361s, respectively.
\item
{\bf Node encode:} Each available node executes {\tt NodeEncode} to compute an
encoded subblock.  Our measurement finds
that the times spent on {\tt NodeEncode} for DRC(9,6,3) and DRC(9,5,3) are
0.067s and 0.068s, which are very similar. Our investigation finds that
DRC(9,6,3) only needs to perform simple node-level encoding, while DRC(9,5,3)
does not even need to perform node-level encoding. Thus, the overhead is
mainly due to the JNI calls rather than the encoding computations.
\item
{\bf Inner-rack transfer:}  We study the inner-rack transfer performance at
the relayer in each non-local rack.  Our measurement using {\tt iperf}
indicates that the effective inner-rack bandwidth of the 10~Gb/s
link is around 9.41~Gb/s~$\approx$~1090~MiB/s.  For DRC(9,6,3), the relayer
receives an amount of $\frac{2}{3}\times 63$~$=$~42~MiB of inner-rack traffic,
so the inner-rack transfer time is 0.039s.  For DRC(9,5,3), the relayer
receives an amount of 64~MiB of inner-rack traffic, so the inner-rack transfer
time is 0.059s.
\item
{\bf Relayer encode:}  Each relayer executes {\tt RelayerEncode}
to re-encode the received encoded subblocks. Our
measurement finds that the times spent on {\tt RelayerEncode} for DRC(9,6,3)
and DRC(9,5,3) are 0.191s and 0.145s, respectively.  Although the relayer
processes more input data in DRC(9,5,3) than in DRC(9,6,3), it performs
simpler linear combinations (see Figure~\ref{fig:963953}) and hence spends
less time in {\tt RelayerEncode}.
\item
{\bf Cross-rack transfer:}  We study the cross-rack transfer performance from
the target's perspective.  Our measurement using {\tt iperf}
indicates that when we set the gateway bandwidth
(i.e., the simulated cross-rack bandwidth) to 1~Gb/s, the effective bandwidth
is around 953~Mb/s~$\approx$~114~MiB/s.  For DRC(9,6,3), the amount of
cross-rack traffic is $2\times63$~$=$~126~MiB, so the cross-rack transfer time
is 1.105s.  For DRC(9,5,3), the amount of cross-rack traffic is 64~MiB, so the
cross-rack transfer time is 0.561s.
\item
{\bf Decode:}  The target executes {\tt Decode} to obtain the reconstructed
block. Our measurement indicates that the times spent on {\tt Decode} for
DRC(9,6,3) and DRC(9,5,3) are 0.443s and 0.32s, respectively.
\end{itemize}

\begin{table}[t]
\centering
\caption{Time breakdown of repairing a single failed block (in seconds).}
\label{tab:break}
\small
\begin{tabular}{|l|c|c|}
\hline 				
& {\bf DRC(9,6,3)} 	& {\bf DRC(9,5,3)} 	\\
\hline
\hline
Disk read		& 0.354			& 0.361			\\
\hline
Node encode		& 0.067			& 0.068			\\
\hline
Inner-rack transfer	& 0.039		& 0.059		\\
\hline
Relayer encode		& 0.191			& 0.145			\\
\hline
Cross-rack transfer	& 1.105		& 0.561		\\
\hline
Decode			& 0.443			& 0.32		\\
\hline
\end{tabular}
\vspace{-6pt}
\end{table}

Table~\ref{tab:break} summarizes the breakdown results.  Our study shows that
the cross-rack transfer time is the most dominant factor in the repair
operation.  If we pipeline all the steps and run them in parallel, we expect
that the repair performance is bottlenecked by the cross-rack transfer.  Note
that the decode time is high in both codes.  Nevertheless, the actual decoding
overhead can be mitigated in node recovery as we can parallelize the decoding
of multiple stripes across different targets.  On the other hand, the disk
read is another dominant factor in the repair performance, especially when the
available cross-rack bandwidth increases.  Our later experiments will further
validate our microbenchmark evaluation.

\subsection{Node Recovery}
\label{subsec:recovery}

We first evaluate the node recovery performance of DoubleR when it repairs
multiple failed blocks of a single failed node.
Specifically, we write 20 stripes of blocks across DataNodes.
To mimic a node failure, we pick one DataNode at random, erase all its 20
blocks, and run DoubleR to repair all erased blocks.  We use the default
parameters (see \S\ref{subsec:methodology}) and vary the gateway bandwidth
from 200~Mb/s to 2~Gb/s.  For each erasure code, we measure the
{\em recovery throughput}, defined as the total size of failed blocks being
repaired divided by the total time of the entire node recovery operation.

\begin{figure}[!t]
\centering
\begin{tabular}{ccc}
\includegraphics[width=1.7in]{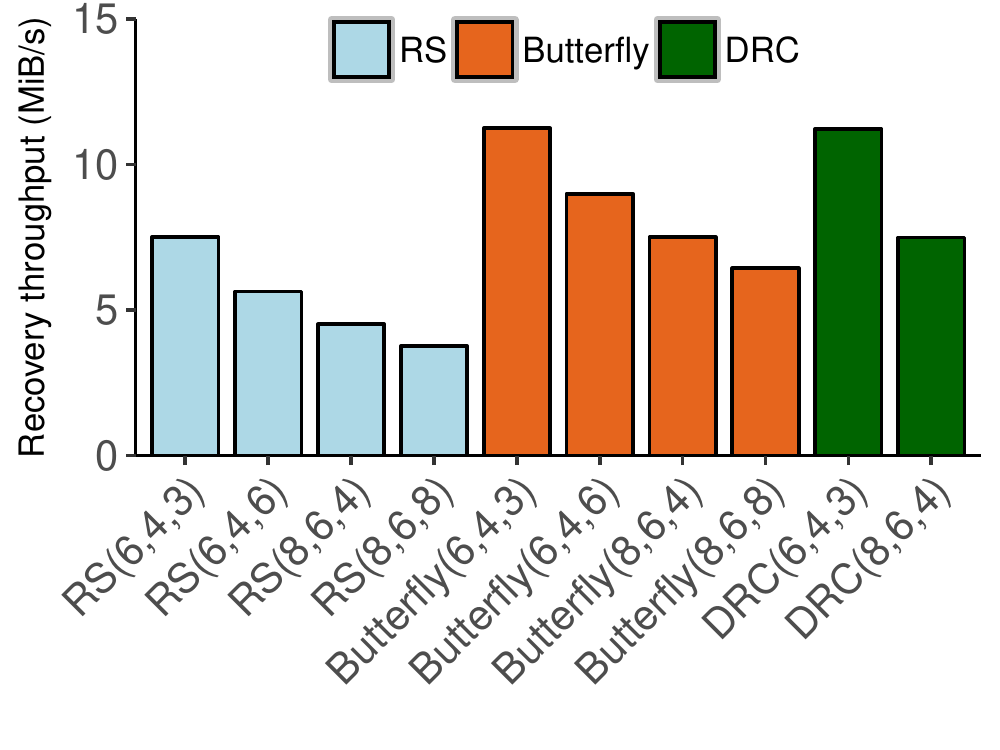} &
\includegraphics[width=1.7in]{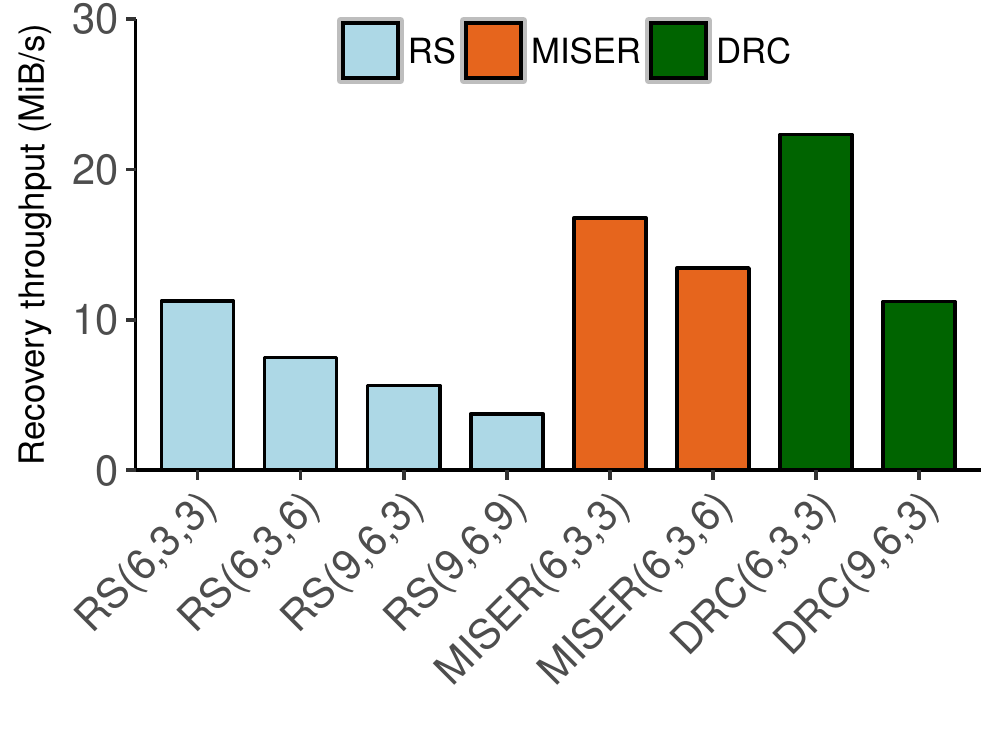} &
\includegraphics[width=1.7in]{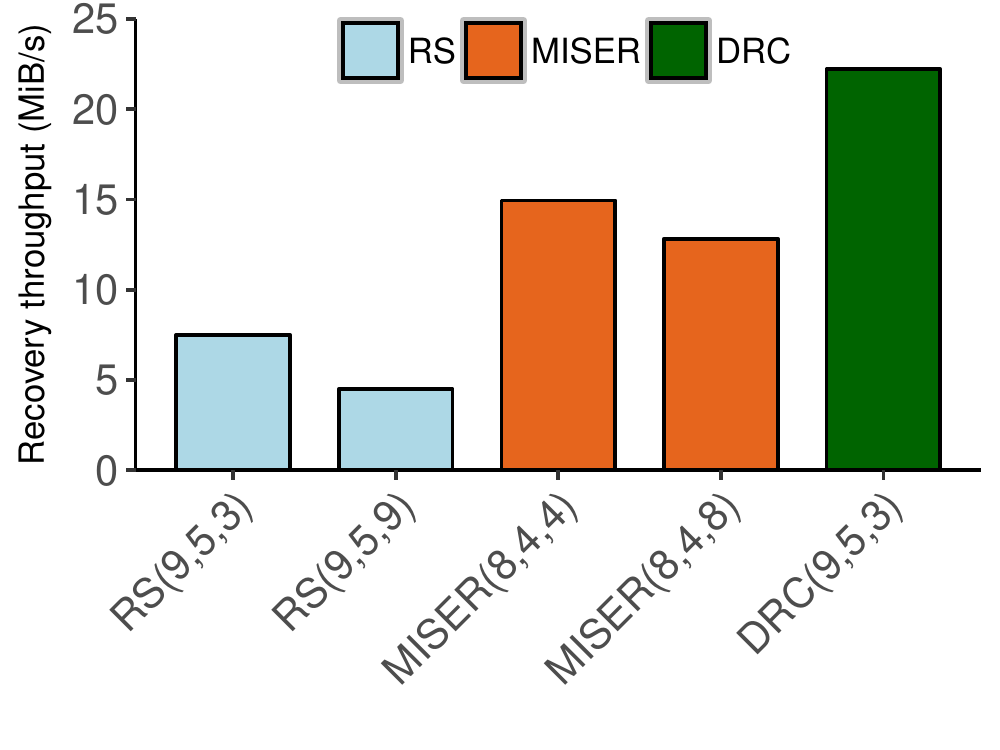}
\vspace{-9pt}\\
\mbox{\small (a) 200~Mb/s, $n-k=2$} &
\mbox{\small (b) 200~Mb/s, $n-k=3$} &
\mbox{\small (c) 200~Mb/s, $n-k=4$}
\vspace{3pt}\\
\includegraphics[width=1.7in]{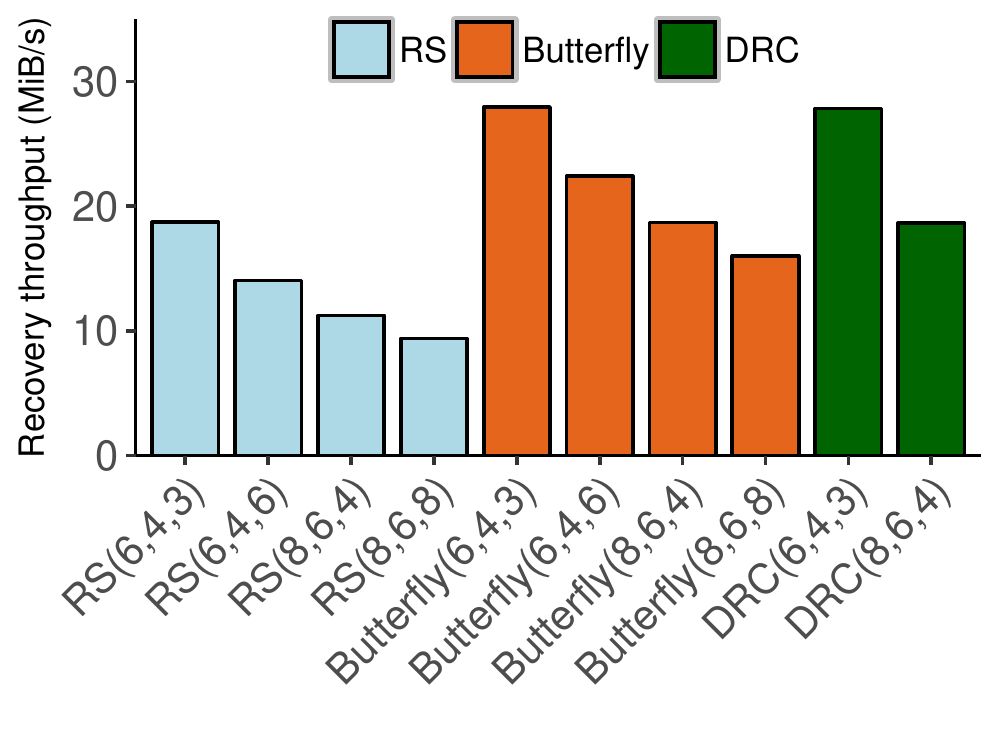} &
\includegraphics[width=1.7in]{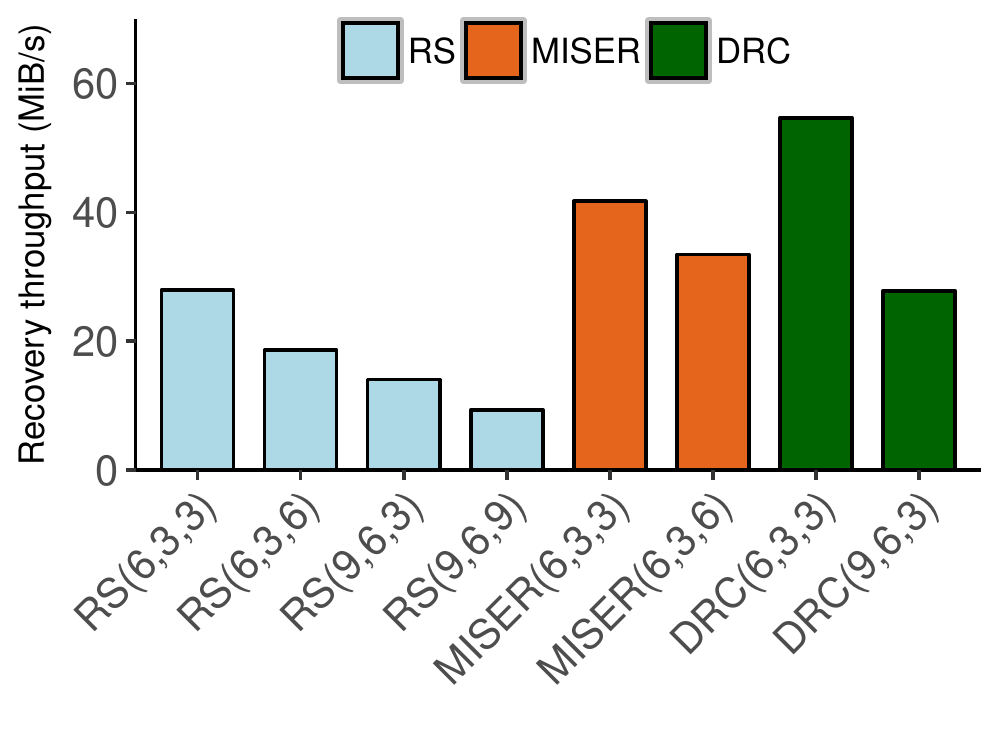} &
\includegraphics[width=1.7in]{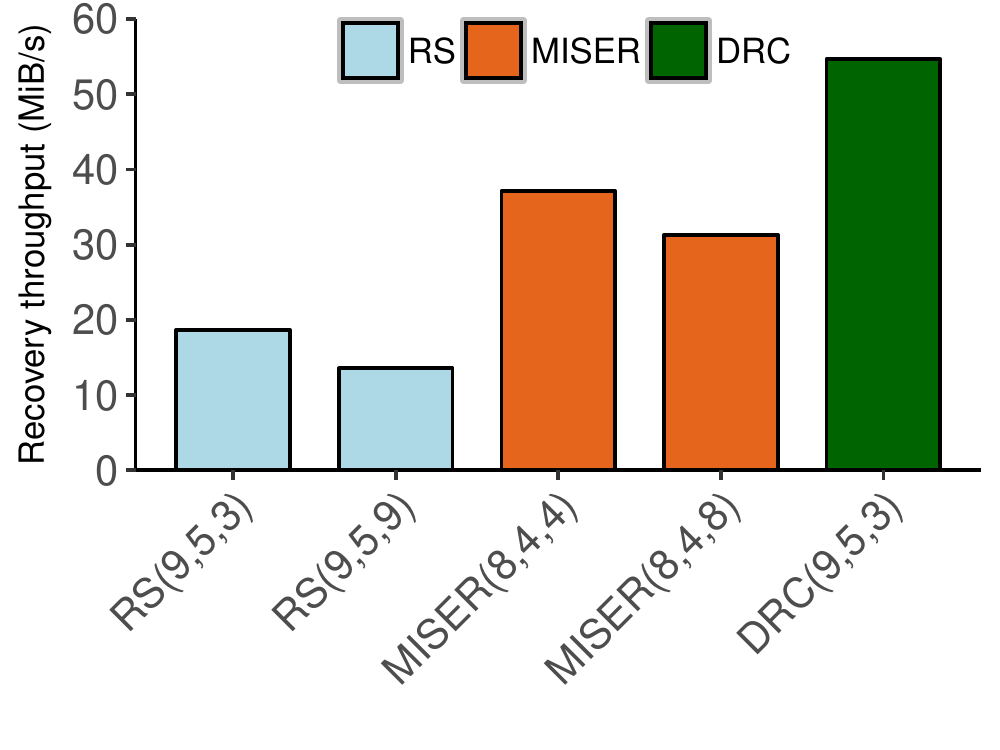}
\vspace{-9pt}\\
\mbox{\small (d) 500~Mb/s, $n-k=2$} &
\mbox{\small (e) 500~Mb/s, $n-k=3$} &
\mbox{\small (f) 500~Mb/s, $n-k=4$}
\vspace{3pt}\\
\includegraphics[width=1.7in]{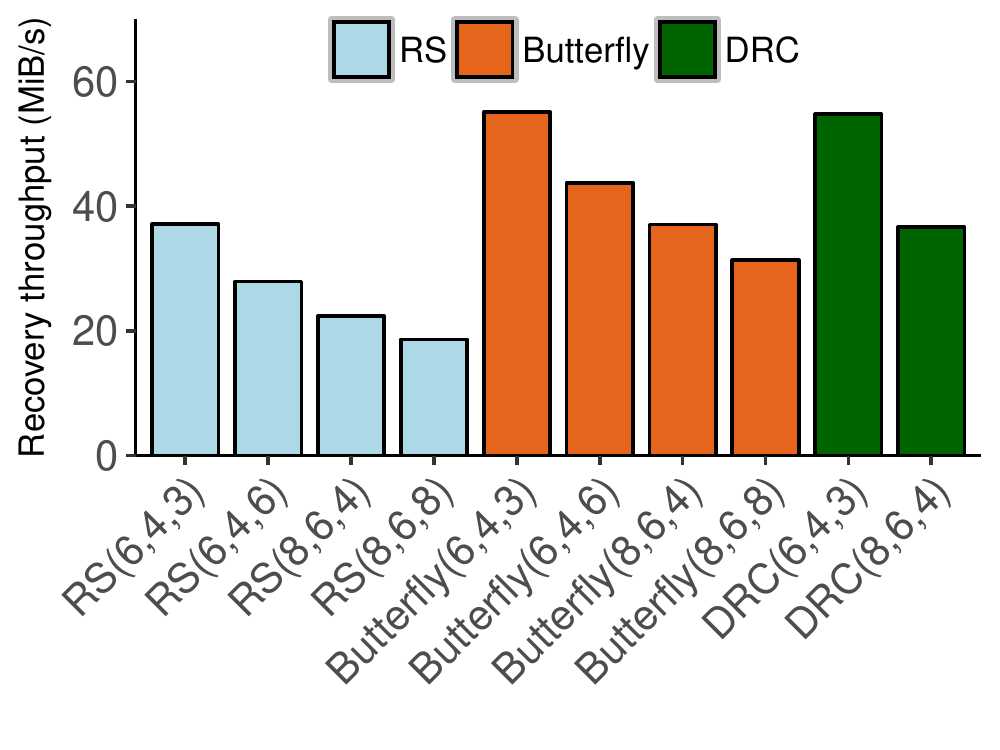} &
\includegraphics[width=1.7in]{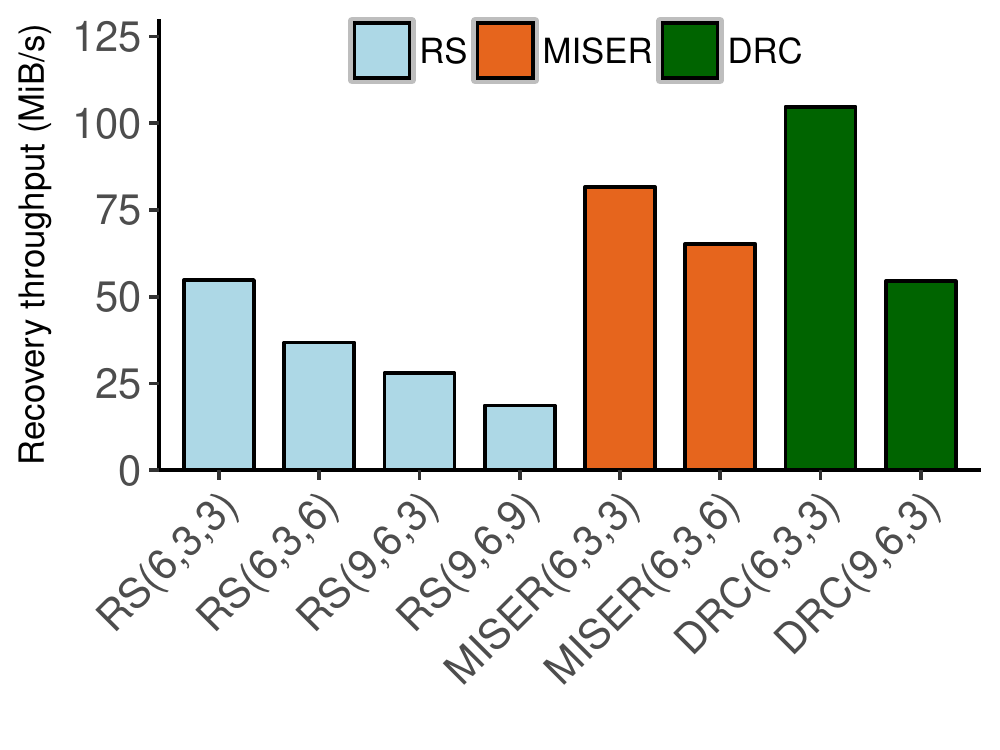} &
\includegraphics[width=1.7in]{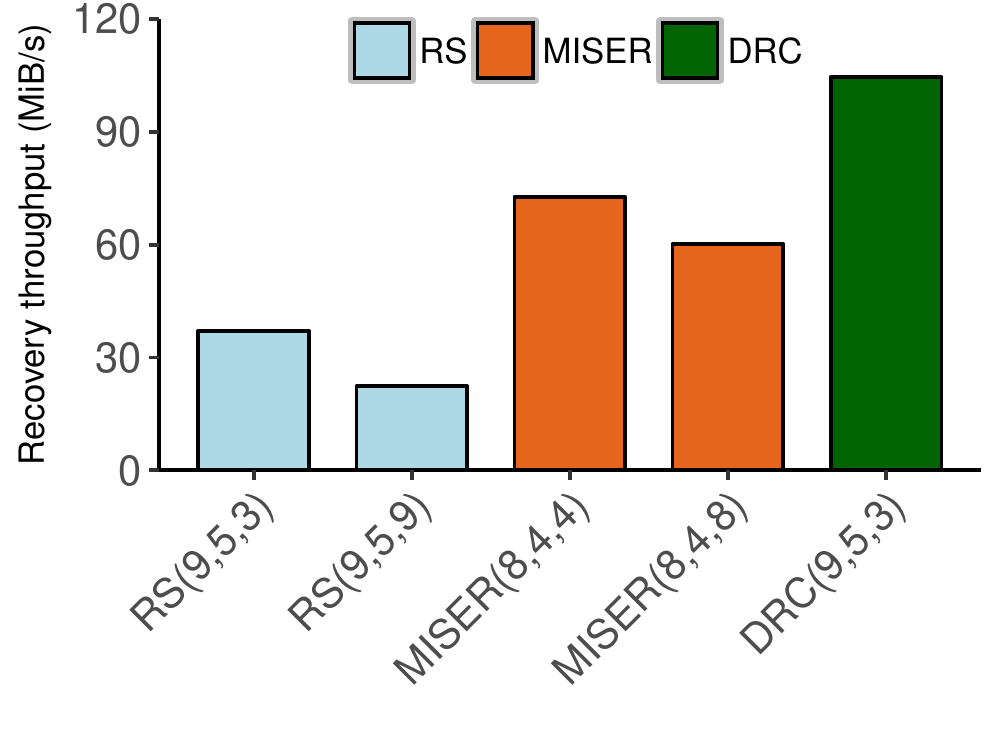}
\vspace{-9pt}\\
\mbox{\small (g) 1~Gb/s, $n-k=2$} &
\mbox{\small (h) 1~Gb/s, $n-k=3$} &
\mbox{\small (i) 1~Gb/s, $n-k=4$}
\vspace{3pt}\\
\includegraphics[width=1.7in]{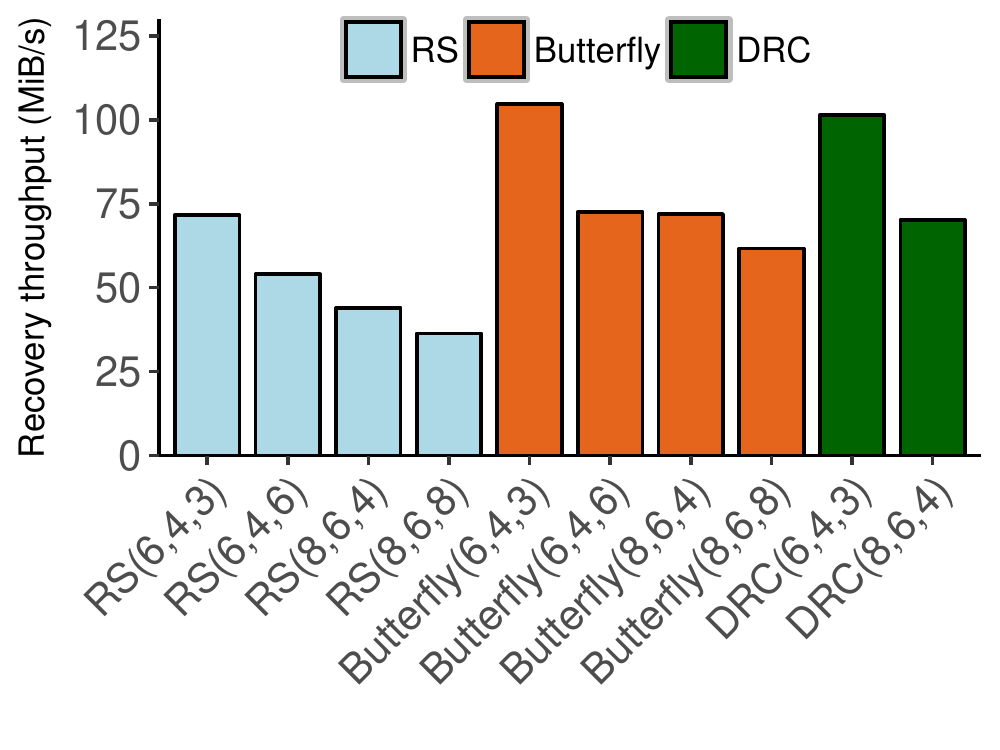} &
\includegraphics[width=1.7in]{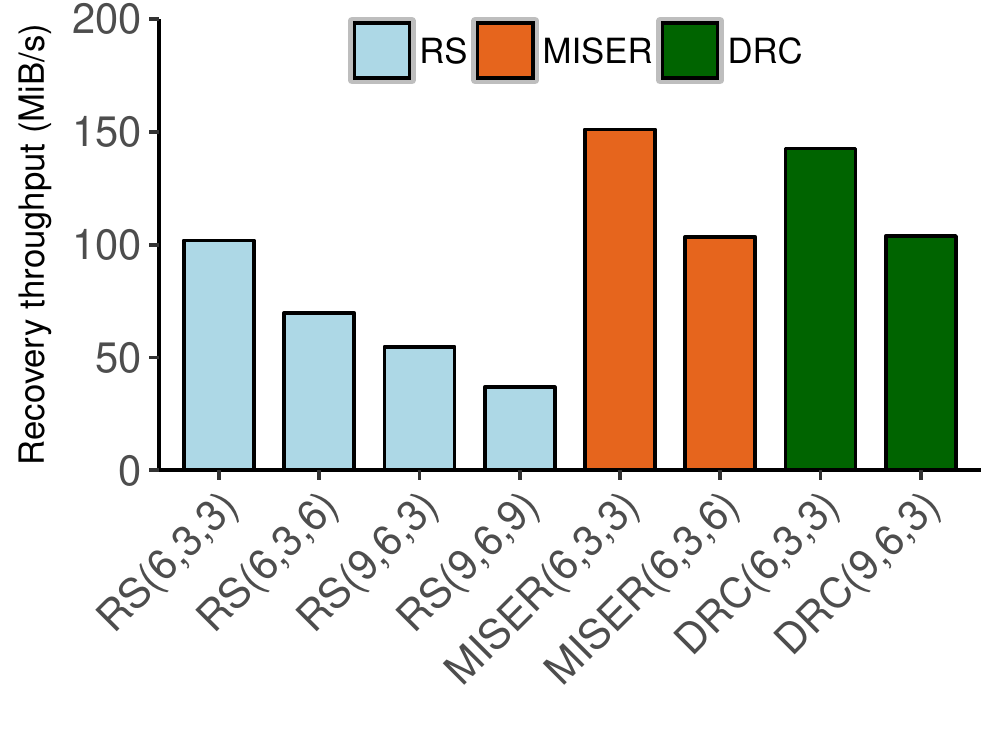} &
\includegraphics[width=1.7in]{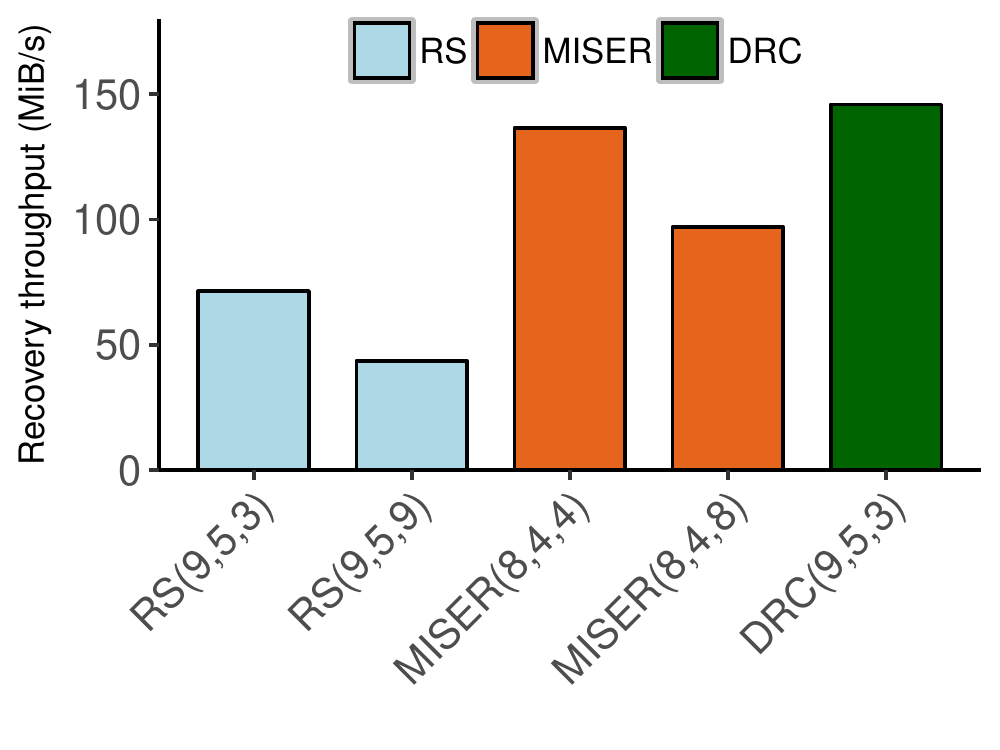}
\vspace{-9pt}\\
\mbox{\small (j) 2~Gb/s, $n-k=2$} &
\mbox{\small (k) 2~Gb/s, $n-k=3$} &
\mbox{\small (l) 2~Gb/s, $n-k=4$}
\end{tabular}
\vspace{-6pt}
\caption{Node recovery performance of different erasure codes under different
gateway bandwidth settings. Note that Butterfly codes and DRC have very close
performance when $r=n/2$ and $n-k=2$ (see Theorem~\ref{theorem:msr}).}
\label{fig:recovery_throughput}
\vspace{-6pt}
\end{figure}

Figure~\ref{fig:recovery_throughput} shows the results for different erasure
codes.   When the gateway bandwidth ranges from 200~Mb/s to 1~Gb/s, the
measured recovery throughput results are fairly consistent with the
numerical results in Figure~\ref{fig:numerical} in \S\ref{subsec:numerical},
as the repair performance is now bottlenecked by the available gateway
bandwidth.  For example, we compare RS(9,5,3) and DRC(9,5,3).  From the
numerical results (see Figure~\ref{fig:numerical}), the cross-rack repair
traffic of RS(9,5,3) is three blocks for repairing a single failed block,
while that of DRC(9,5,3) is one block only.
From the measured results, the recovery throughput of DRC(9,5,3) is
2.96$\times$, 2.92$\times$, and 2.81$\times$ that of RS(9,5,3) when the
gateway bandwidth is 200~Mb/s, 500~Mb/s, and 1~Gb/s, respectively (see
Figures~\ref{fig:recovery_throughput}(c), \ref{fig:recovery_throughput}(f),
and \ref{fig:recovery_throughput}(i), respectively).  Overall, when the
available gateway bandwidth is smaller, the recovery throughput gain is closer
to the theoretical gain.

On the other hand, when the gateway bandwidth is 2~Gb/s, the disk read also
becomes a dominant factor in the repair performance (see
\S\ref{subsec:microbenchmarks}), so the gain of DRC diminishes.  For example,
the recovery throughput gain of DRC(9,5,3) over RS(9,5,3) drops to
2.04$\times$ (see Figure~\ref{fig:recovery_throughput}(l)).  In another
example, DRC(6,3,3) has 10\% lower recovery throughput than MISER(6,3,3) when
the gateway bandwidth is 2~Gb/s (see Figure~\ref{fig:recovery_throughput}(k)),
although it has higher throughput than MISER(6,3,3) when the gateway bandwidth
is no more than 1~Gb/s.  Thus, we can claim the benefits of DRC only if
cross-rack transfer is the performance bottleneck in a data center.

\subsection{Degraded Reads}

\begin{figure}[t]
\centering
\begin{tabular}{ccc}
\includegraphics[width=1.7in]{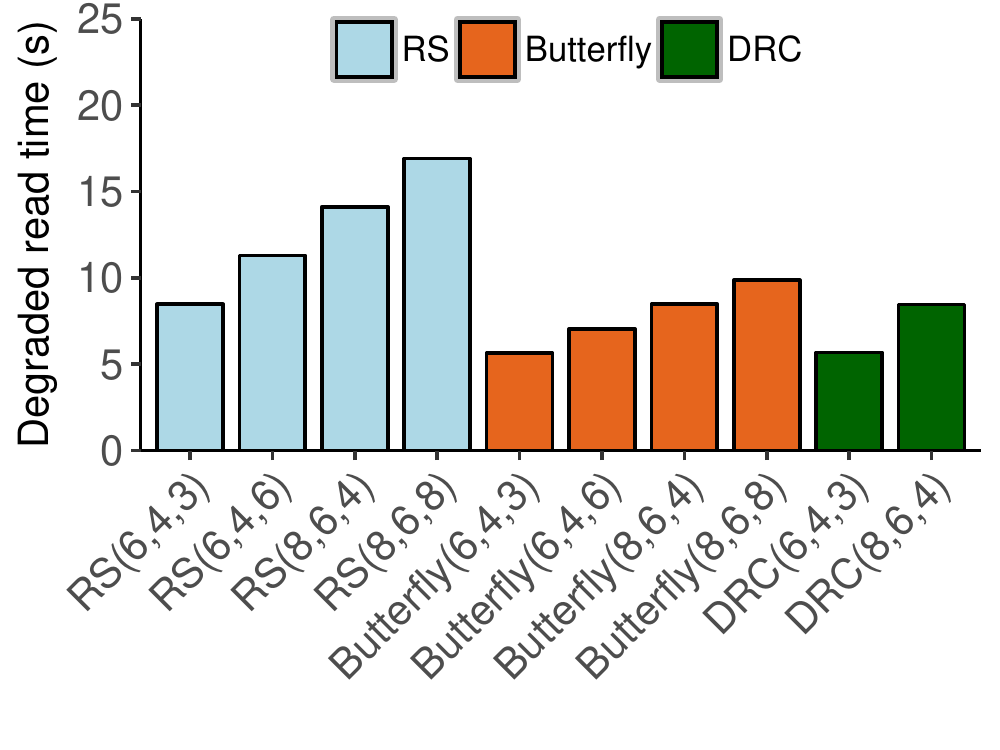} &
\includegraphics[width=1.7in]{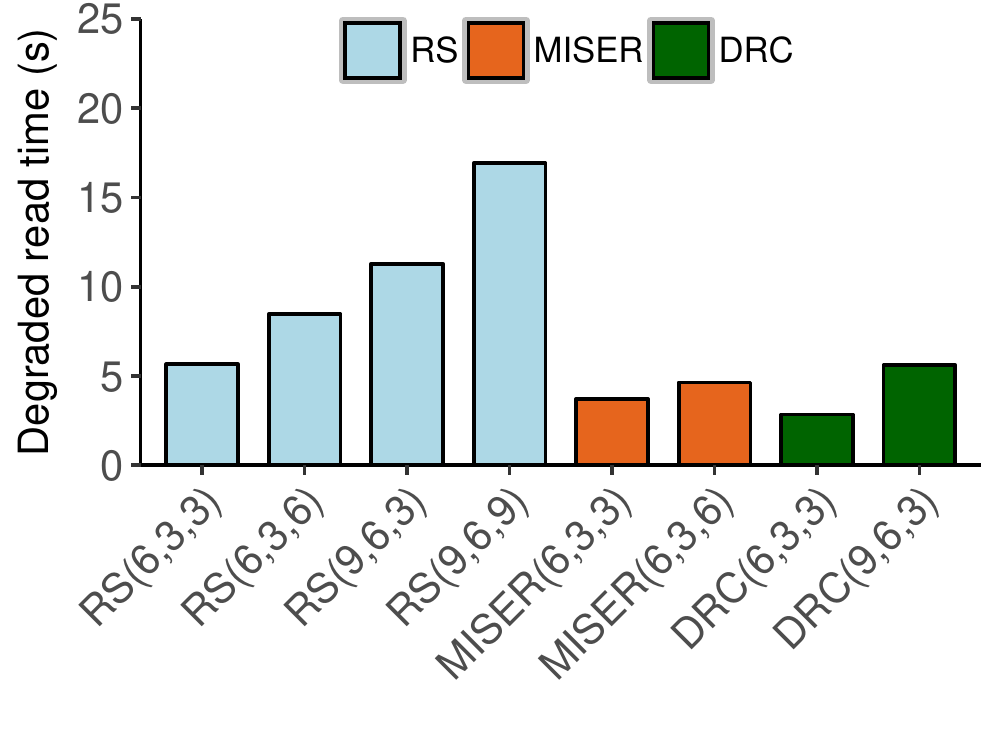} &
\includegraphics[width=1.7in]{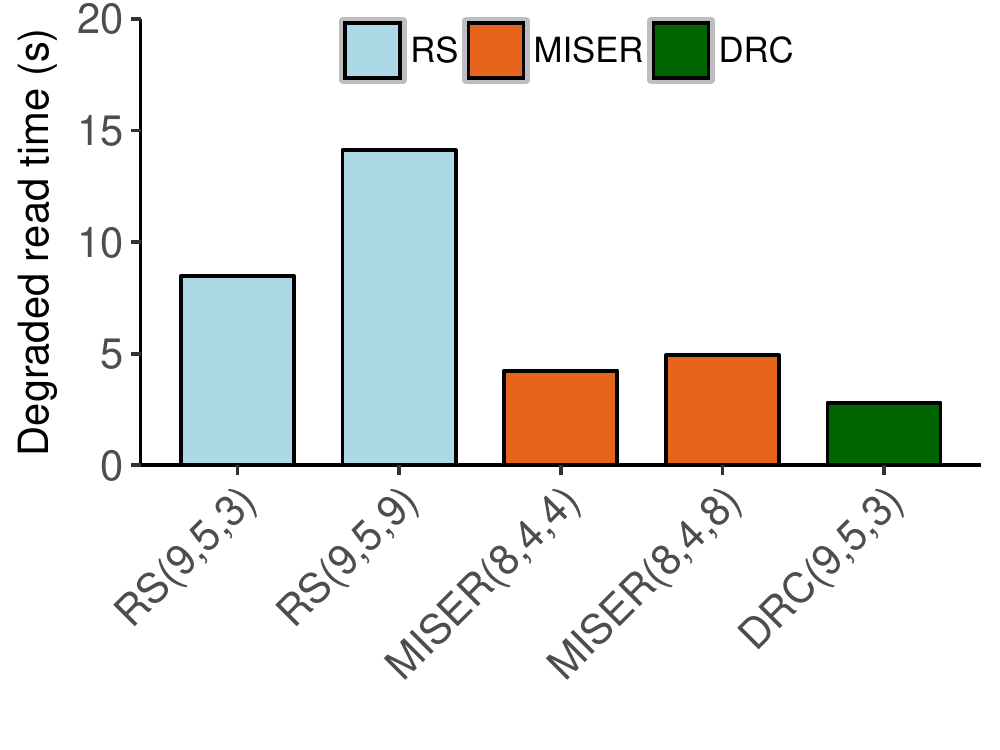}
\vspace{-9pt}\\
\mbox{\small (a) 200~Mb/s, $n-k=2$} &
\mbox{\small (b) 200~Mb/s, $n-k=3$} &
\mbox{\small (c) 200~Mb/s, $n-k=4$}
\vspace{3pt}\\
\includegraphics[width=1.7in]{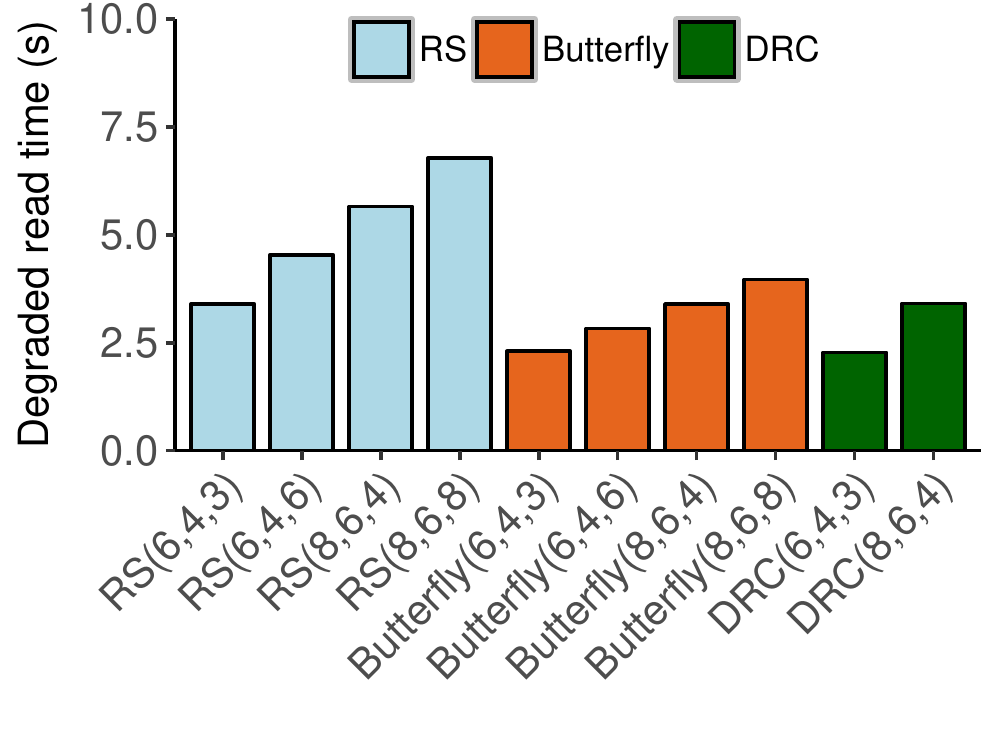} &
\includegraphics[width=1.7in]{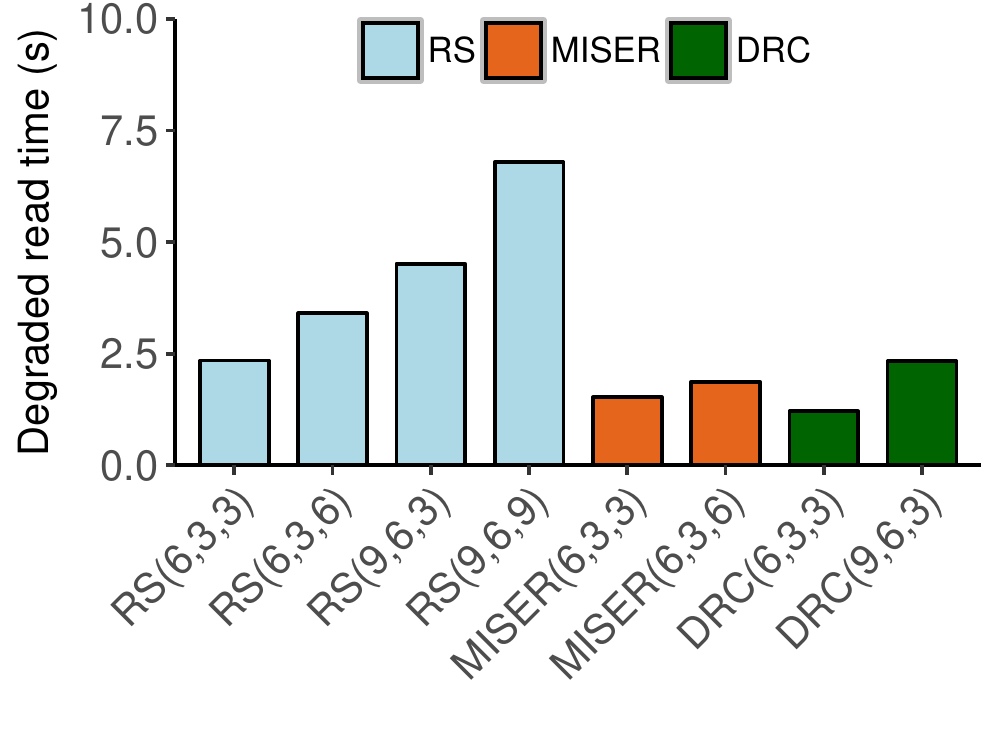} &
\includegraphics[width=1.7in]{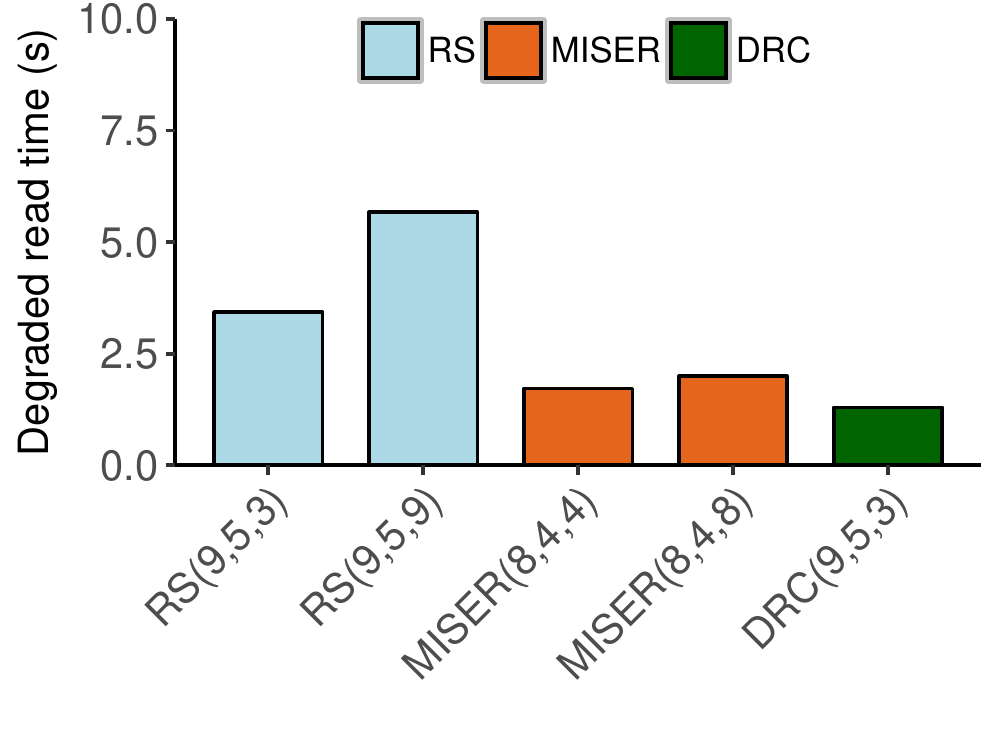}
\vspace{-9pt}\\
\mbox{\small (d) 500~Mb/s, $n-k=2$} &
\mbox{\small (e) 500~Mb/s, $n-k=3$} &
\mbox{\small (f) 500~Mb/s, $n-k=4$}
\vspace{3pt}\\
\includegraphics[width=1.7in]{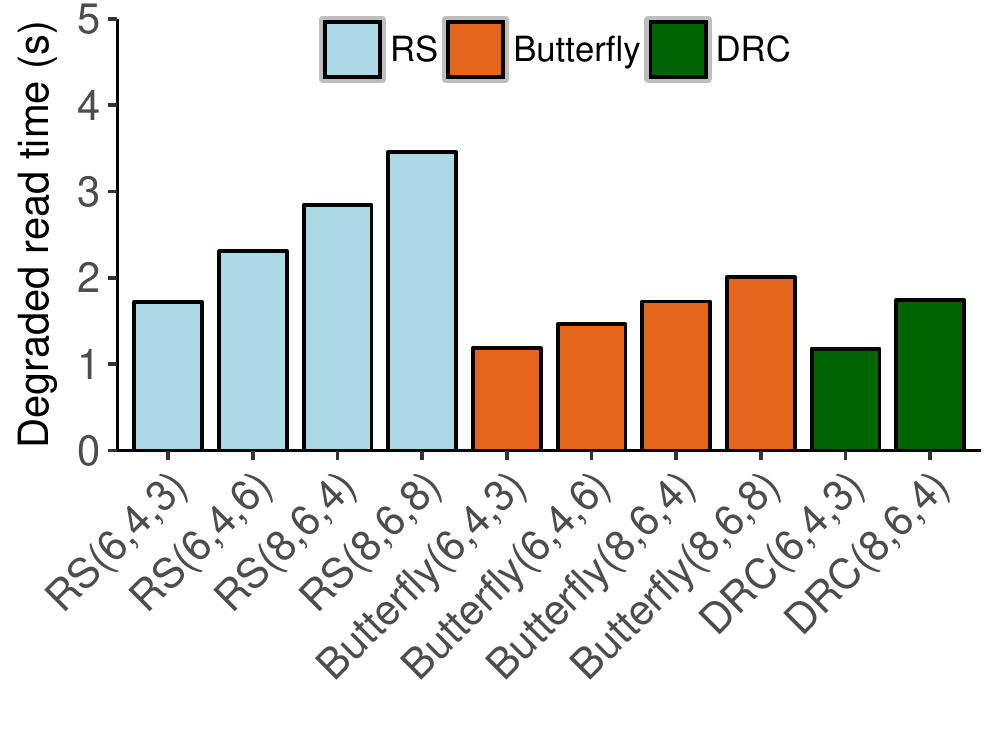} &
\includegraphics[width=1.7in]{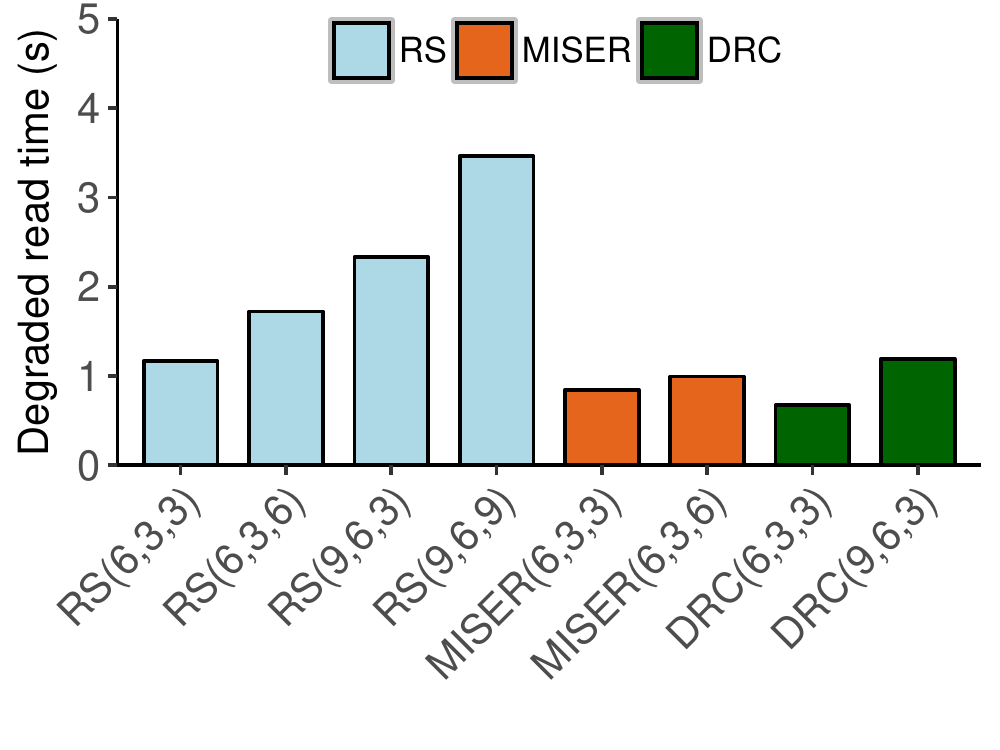} &
\includegraphics[width=1.7in]{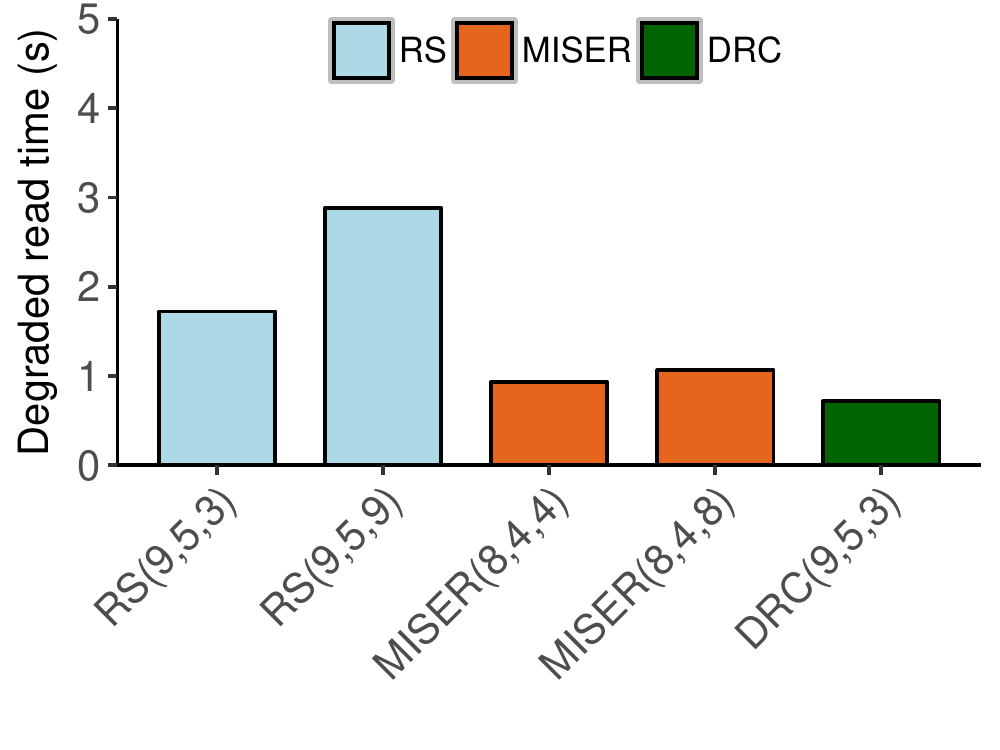}
\vspace{-9pt}\\
\mbox{\small (g) 1~Gb/s, $n-k=2$} &
\mbox{\small (h) 1~Gb/s, $n-k=3$} &
\mbox{\small (i) 1~Gb/s, $n-k=4$}
\vspace{3pt}\\
\includegraphics[width=1.7in]{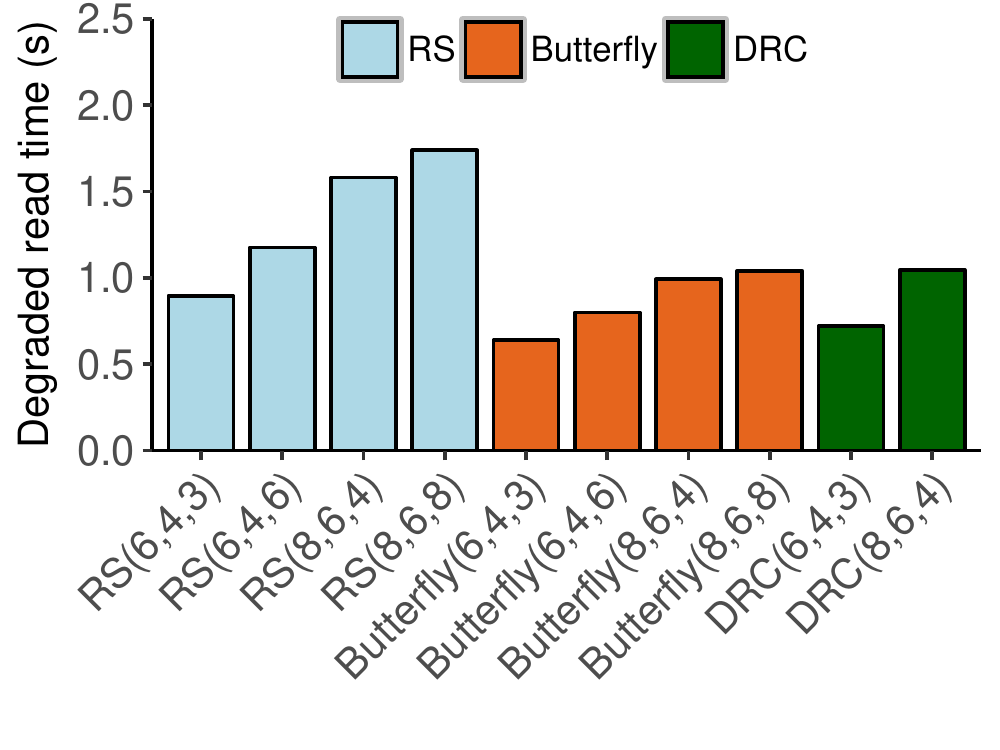} &
\includegraphics[width=1.7in]{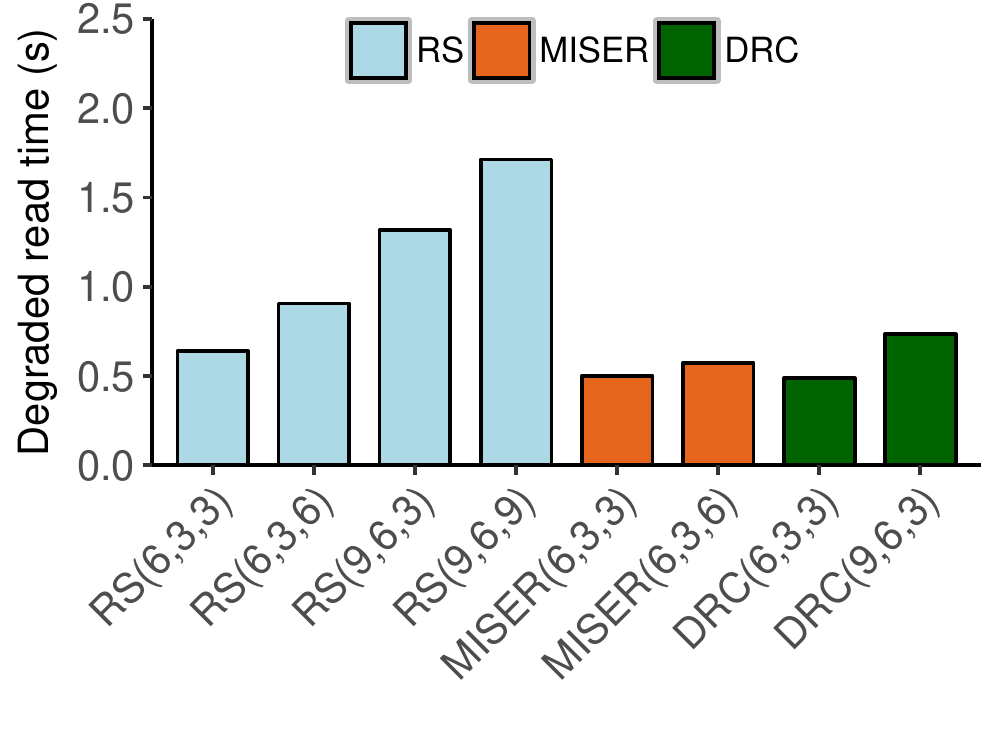} &
\includegraphics[width=1.7in]{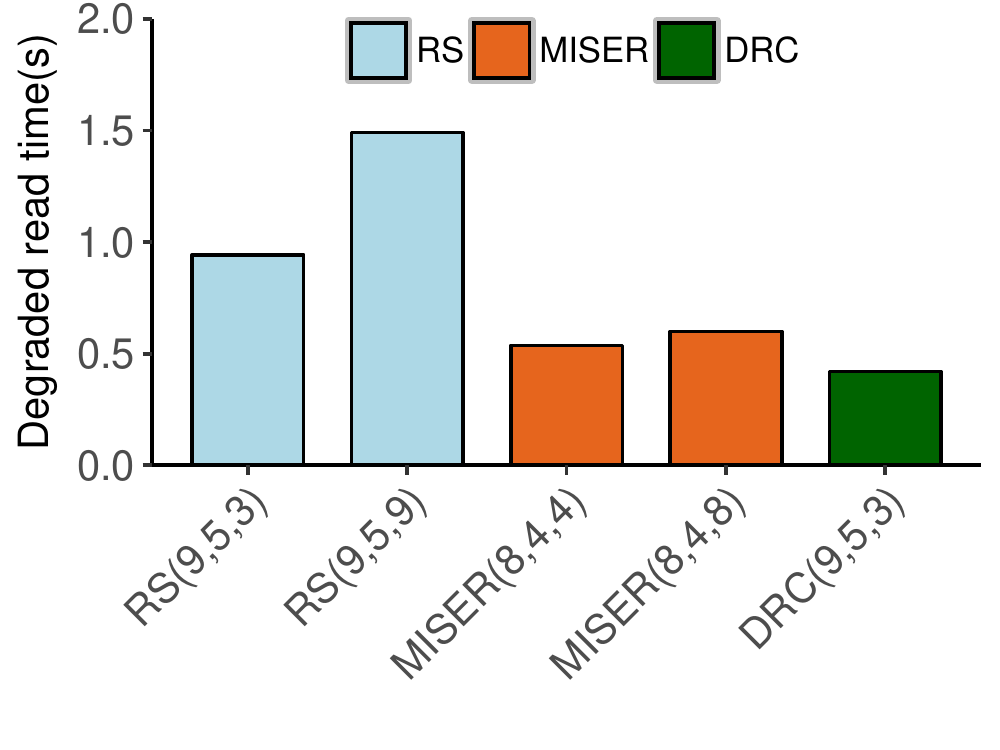}
\vspace{-9pt}\\
\mbox{\small (j) 2~Gb/s, $n-k=2$} &
\mbox{\small (k) 2~Gb/s, $n-k=3$} &
\mbox{\small (l) 2~Gb/s, $n-k=4$}
\end{tabular}
\vspace{-6pt}
\caption{Degraded read performance of different erasure codes under different
gateway bandwidth settings.}
\label{fig:degraded}
\vspace{-6pt}
\end{figure}

We next evaluate the degraded read performance when the file system client
accesses a single unavailable block.
Specifically, we randomly choose a data block to erase and let the file system
client access the erased block through a degraded read.  As in
\S\ref{subsec:recovery}, we again use the default parameters and vary the
gateway bandwidth from 200~Mb/s to 2~Gb/s.  We measure the {\em degraded read
time}, defined as the latency from issuing a read request until the failed
block is completely reconstructed at the file system client.

Figure~\ref{fig:degraded} shows the results for different erasure codes. We
see that DRC also shows performance gain in degraded reads by minimizing the
cross-rack repair traffic.
For example, we compare RS(9,5,3) and DRC(9,5,3).  The degraded read time of
DRC(9,5,3) is 66.9\%, 62.3\%, 58.0\%, and 55.4\% less than that of RS(9,5,3)
when the gateway bandwidth is set to 200~Mb/s, 500~Mb/s, 1~Gb/s, and 2~Gb/s,
respectively.

\subsection{Impact of Strip Size and Block Size}

We finally evaluate the repair performance of DoubleR for various strip sizes
and block sizes.  Here, we focus on node recovery as in
\S\ref{subsec:recovery}, and compare DRC(6,4,3), DRC(6,3,3), DRC(8,6,4),
and DRC(9,5,3).  Note that we do not consider DRC(9,6,3) as its strip size and
block size are different from others (see \S\ref{subsec:methodology}).  We set
the gateway bandwidth as 1~Gb/s.

\begin{figure}[t]
\centering
\begin{tabular}{cc}
\includegraphics[width=2.5in]{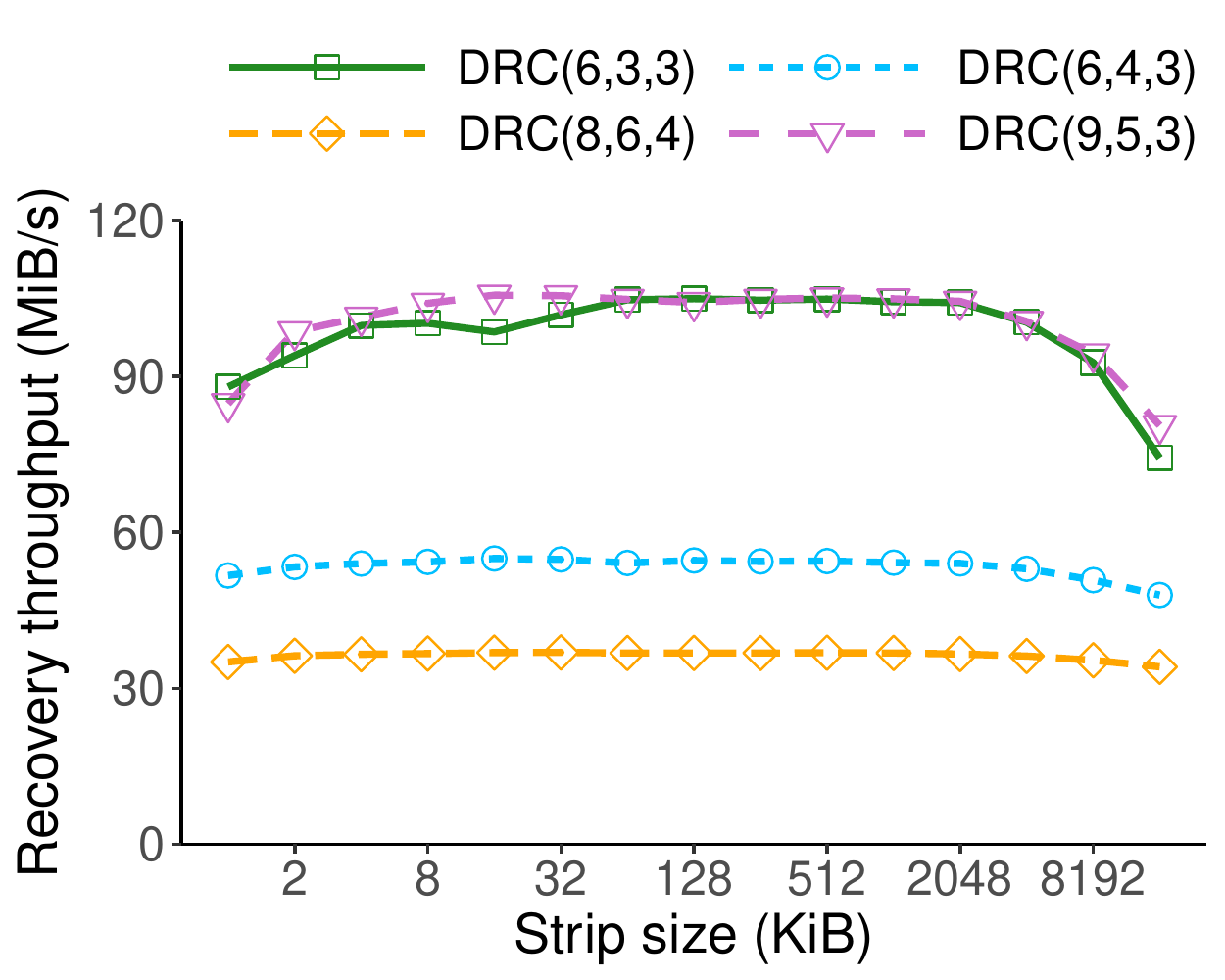} &
\includegraphics[width=2.5in]{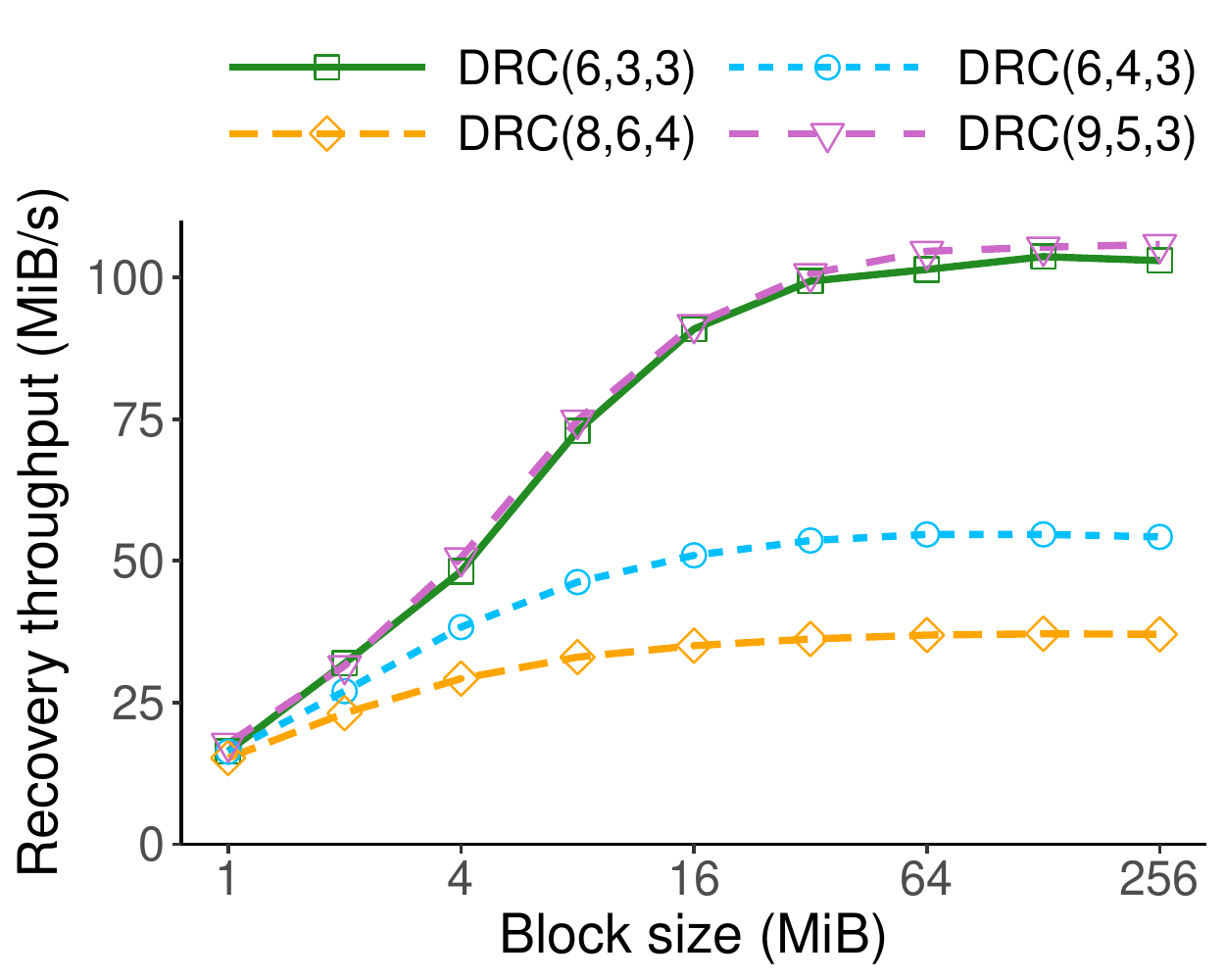} \\
\mbox{\small (a) Strip size} &
\mbox{\small (b) Block size}
\end{tabular}
\vspace{-6pt}
\caption{Impact of strip size and block size on node recovery performance.}
\label{fig:sizetest}
\end{figure}

Figure~\ref{fig:sizetest}(a) first shows the recovery throughput when the
strip size varies from 1~KiB to 16~MiB, where the block size is fixed at
64~MiB.  We see that there is a performance drop when the strip size is too
small or too large.  Specifically, when the strip size is less than 8~KiB,
DoubleR needs to issue more function calls to access more strips of a block,
and the overhead becomes more significant. When the strip size is larger than
2~MiB, the parallelism across multiple strips of a block cannot be fully
utilized.  The recovery throughput is the maximum when the strip size is in
between.

Figure~\ref{fig:sizetest}(b) shows the recovery throughput when the block size
varies from 1~MiB to 256~MiB, where the strip size is fixed at 256~KiB.  The
recovery throughput is small when the block size is small, as the block access
overhead is significant.  The recovery throughput increases with the block
size and reaches the maximum when the block size is at least 64~MiB.

\section{Related Work}
\label{sec:related}

We review related work on erasure coding in the context of improving repair
performance. 

\paragraph{Erasure code constructions:}  Many constructions of erasure codes
have been proposed to reduce the repair traffic.  Regenerating codes
\cite{dimakis10} are a special family of erasure codes that minimize the
repair traffic and provably achieve the optimal trade-off between storage
redundancy and repair traffic.  Constructions of regenerating codes have been
proposed, such as Interference Alignment codes \cite{wu09b,suh11b,shah12a},
Product-Matrix codes \cite{rashmi11}, Zigzag codes \cite{tamo13}, FMSR codes
\cite{chen14,hu17}, PM-RBT codes \cite{rashmi15}, and Butterfly codes
\cite{pamies16}.  As stated in \S\ref{subsec:goals}, recent studies 
\cite{ye17a,ye17b,goparaju17,sasidharan16} also propose MSR code constructions
for general parameters.

Some erasure codes aim to minimize I/O (i.e., the amount of data read from
storage) during repair.  For example, Rotated RS codes \cite{khan12} and
Hitchhiker \cite{rashmi14} propose new parity constructions that send fewer
blocks in a single-node failure repair. 

Some erasure codes trade storage efficiency for repair performance. 
Simple Regenerating Codes (SRC) \cite{papailiopoulos12} retrieve
data from a small number of non-failed nodes to repair a failed node, thereby
limiting the I/O overhead of accessing non-failed nodes during repair. 
Locally repairable codes trade storage efficiency for repair performance by
associating local parity blocks with different subsets of nodes.  Thus, they
can retrieve data from a smaller number of nodes during repair, and limit both 
repair traffic and I/O.  Two representative constructions are Azure's LRC
\cite{huang12} and Facebook's LRC \cite{sathiamoorthy13}.  

The above erasure codes mainly adopt flat block placement in hierarchical
data centers to tolerate rack failures (as mentioned in
\cite{ford10,huang12,sathiamoorthy13,muralidhar14,rashmi14,rashmi15}).  Our
work complements the above studies by specifically minimizing the critical
cross-rack repair traffic via hierarchical block placement.

\paragraph{Erasure coding in hierarchical data centers:} Erasure-coded repair
in hierarchical data centers has been studied, but in a limited context.  Some
studies focus on a data center with two racks \cite{gaston13,pernas13}, or
propose locally repairable codes for multiple racks \cite{tebbi14}.  R-STAIR
codes \cite{li16} place an extra parity block in each rack to allow rack-local
repair without cross-rack traffic.  However, R-STAIR codes require
sophisticated configurations of parameters of full-rack and partial-rack fault
tolerance.  CAR \cite{shen16} is specifically designed for RS codes, and
exploits intra-rack encoding to reduce the cross-rack repair traffic.
However, CAR does not provide any theoretical guarantee of minimizing the
cross-rack repair traffic as in DRC \cite{hu16}.  We extend the DRC
framework \cite{hu16} from the applied side: we provide practical DRC
constructions and evaluate their prototype implementation. 

\paragraph{Efficient repair approaches:}  Some studies propose efficient
repair approaches for existing erasure codes.  For example, lazy repair
\cite{bhagwan04,silberstein14} triggers repair only when the number of failures
reaches a threshold to avoid repairing temporary failures.  CORE \cite{li15}
extends existing regenerating codes to support the optimal recovery of
multi-node failures, and presents a prototype implementation on HDFS.
HACFS \cite{xia15} dynamically switches encoded blocks between two erasure
codes to balance storage overhead and repair performance.  PPR \cite{mitra16}
divides a repair into partial operations executed by multiple servers in
parallel to reduce the overall repair time.  
Repair pipelining \cite{li17} further reduces the repair time to almost the
same as the normal read time by slicing the repair along a linear chain.  
Our work differs from them by proposing new regenerating code constructions
for hierarchical data centers. 

\section{Discussion}
\label{sec:discuss}

The repair gains of DRC build on several design assumptions.  In this section,
we discuss the design trade-offs of DRC.

\paragraph{Reduced rack-level fault tolerance:}  DRC builds on hierarchical
block placement to trade reduced rack-level fault tolerance for the
minimum cross-rack repair traffic.  The underlying assumption is that
rack failures or correlated node failures are rare, so minimizing the
cross-rack repair traffic can improve the repair performance and hence the
overall storage reliability (see \S\ref{subsec:reliability}).  Otherwise,
erasure codes that build on flat block placement should be used. 

\paragraph{Limited cross-rack bandwidth:}  DRC assumes that the repair
performance is bottlenecked by the constrained cross-rack bandwidth.  If the
cross-rack bandwidth is sufficient, then other types of overhead may become
prohibitive.  For example, Family~1 of DRC needs to read $n-1$ blocks from
disk to achieve the minimum cross-rack repair traffic (same for MSR codes),
while RS codes only need to read $k$ blocks.  Also, while DRC minimizes the
cross-rack repair traffic, its total number of blocks being transferred,
including both cross-rack and inner-rack, is more than that of MSR codes in
general (e.g., see Figure~\ref{fig:motivating} in \S\ref{subsec:examples}).
Thus, the repair gain of DRC may no longer hold when the cross-rack bandwidth
is sufficient, as shown in \S\ref{subsec:recovery} for the case of 2~Gb/s
gateway bandwidth. 

\paragraph{Limited parameters:}  Our current DRC constructions are designed
for specific sets of parameters.  An open question is whether there exist
explicit DRC constructions for general sets of parameters. 

\paragraph{Storage optimality:}  In this paper, we only focus on erasure codes
that are MDS, including RS codes, MSR codes, and DRC.  On the other hand, if
we relax the storage optimality assumption, we can further reduce or even
eliminate the cross-rack repair traffic.  For example, locally repairable
codes \cite{huang12,sathiamoorthy13} can be deployed via hierarchical block
placement by placing each local parity stripe in the same rack, so as to
eliminate the cross-rack repair traffic in a single-node repair.  The
trade-off of locally repairable codes is that they are non-MDS and incur
higher storage redundancy than MDS codes. 

\section{Conclusions}
\label{sec:conclusions}

We present DoubleR, a framework that realizes repair layering to improve
repair performance in hierarchical data centers.  DoubleR builds on the recent
theoretical findings of DRC and aims to minimize the cross-rack repair
traffic.  We design and implement two families of practical DRC
constructions for DoubleR.  Experiments on our DoubleR prototype show the
effectiveness of DRC in terms of node recovery throughput and degraded read
time over state-of-the-art regenerating codes.  The source code of our DoubleR
prototype is available for download at 
{\bf http://adslab.cse.cuhk.edu.hk/software/doubler}.

\bibliographystyle{abbrv}
\bibliography{paper}

\end{document}